\documentclass[journal]{IEEEtran}

\IEEEoverridecommandlockouts
\usepackage{cite}
\usepackage{lipsum}
\usepackage{diagbox}
\usepackage{graphicx}
\usepackage{caption}
\usepackage{epsfig}
\usepackage{dblfloatfix}
\usepackage{blindtext}
\usepackage{float}
\usepackage[cmex10]{amsmath}
\usepackage{algorithmic}
\usepackage{amssymb}
\usepackage{color}
\usepackage{bm}
\usepackage{amsfonts}
\usepackage{amsmath}
\usepackage{amsmath,mathtools,amssymb}
\usepackage{graphicx}
\usepackage{subfig}
\usepackage{tablefootnote}
\usepackage{placeins}
\usepackage{tikz}
\usetikzlibrary{positioning,shapes}
\usepackage{bookmark}

\DeclarePairedDelimiter{\norm}{\lVert}{\rVert}
\usepackage[mathlines]{lineno}
\usepackage{booktabs,multirow}
\usepackage{balance}
\newcommand{\RomanNumeralCaps}[1]
    {\MakeUppercase{\romannumeral #1}}
\usepackage{tikz}
\usetikzlibrary{shapes}
\newcommand*{\circled}[2][blue]{
  \tikz[baseline=(char.base)]{
              \node[shape=ellipse,inner sep=2pt,
                draw=#1,
             ] (char) {#2};}
}

\newcommand{\bb}[1]{#1}

\begin{document}

\title{Deep Unfolding Basis Pursuit: Improving Sparse Channel Reconstruction via Data-Driven Measurement Matrices}

\author{Pengxia~Wu and
        	Julian~Cheng,~\IEEEmembership{Senior~Member,~IEEE}

\thanks{P. Wu and J. Cheng are with the School of Engineering, The University
of British Columbia, Kelowna, BC V1X 1V7, Canada (e-mail: pengxia.wu@alumni.ubc.ca, julian.cheng@ubc.ca).}
}
\maketitle

\begin{abstract}

 	For massive multiple-input multiple-output (MIMO) systems operating in frequency-division duplex mode, downlink channel state information (CSI) acquisition will incur large overhead. This overhead is substantially reduced when sparse channel estimation techniques are employed, owing to the channel sparsity in the angular domain.  When a sparse channel estimation method is implemented, the measurement matrix, which is related to the pilot matrix, is essential to the channel estimation performance. Existing sparse channel estimation schemes widely adopt random measurement matrices, which have been criticized for their suboptimal reconstruction performance. \bb{This paper proposes novel data-driven solutions to design the measurement matrix. Model-based autoencoders are customized to optimize the measurement matrix by unfolding the classical basis pursuit algorithm. The obtained data-driven measurement matrices are applied to existing sparse reconstruction algorithms, leading to ﬂexible hybrid data-driven implementations for sparse channel estimation. } Numerical results show that the proposed data-driven measurement matrices can achieve more accurate reconstructions and use fewer measurements than the existing random matrices, thereby leading to a higher achievable rate for CSI acquisition.  \bb{Moreover, compared with existing pure deep learning-based sparse reconstruction methods, the proposed hybrid data-driven scheme, which uses the novel data-driven measurement matrices with conventional sparse reconstruction algorithms, can achieve higher reconstruction accuracy.}

\end{abstract}

\IEEEpeerreviewmaketitle
\section{Introduction}
	
	Massive multiple-input multiple-output (MIMO) is a key technology for next-generation wireless communications due to its high capacity and ability to combat the \mbox{small-scale} fading of wireless channels \cite{swindlehurst2014millimeter}. 
	In massive MIMO systems, accurate downlink channel state information (CSI) is essential to obtain expected array gains from \mbox{large-scale} antenna arrays by exploiting transmit beamforming \cite{molisch2017hybrid}. 
	However, it is challenging to acquire downlink CSI for a massive MIMO system operating in frequency-division duplex (FDD) mode.
	An FDD massive MIMO system cannot exploit channel reciprocity since the uplink and downlink channels occupy different spectral bands. 
	Downlink channels are often estimated at the user equipments (UEs) and then the estimated CSI is sent back to the base station (BS).
	This process of downlink CSI acquisition consumes prohibitively high communication resources because the overheads of both downlink pilot transmission and CSI feedback are proportional to the number of BS antennas, which is large in massive MIMO systems.
	To reduce the overheads of downlink CSI acquisition, sparse channel estimation has been introduced.
	By exploiting the channel sparsity and employing compressive sensing techniques, one can use a few pilots to sense high-dimensional channels and then reconstruct the channels accurately \cite{choi2017compressed}. 
	The overheads of downlink pilots and CSI feedback are no longer proportional to the number of BS antennas, but mostly depend on the sparsity level of the channel.
	Thus, the overheads of downlink pilots and CSI feedback can be substantially reduced.
	
	Massive MIMO channels are high-dimensional and exhibit sparsity features in the angular domain (or beamspace) due to the limited number of scattering clusters and the small angular spreads in propagation environments.
 	The majority of channel components are either zero or approximately zero, and only a few components have relatively large magnitudes.
	Based on this beamspace channel sparsity, various beamspace channel estimation algorithms \cite{gao2016structured,rao2014distributed,gao2015spatially, shen2016joint} and CSI feedback reduction schemes \cite{LiuCSI2019, ShenCompressed2016,song2010compressive} use compressive sensing techniques to enable sparse channel estimation. 
	To implement the compressive-sensing-aided sparse channel estimation successfully,  we require three key components. 
	The first component is the sparse channel representation. 
	Massive MIMO channels can be sparse under an appropriate spatial basis.
	The most common sparse representation is the beamspace channel representation, which can be obtained by a discrete Fourier transform (DFT) of the spatial channels \cite{brady2013beamspace}.
	The second component is the sparse recovery algorithm. 
	Based on valid sparse channel representations, one needs an efficient sparse recovery algorithm to reconstruct unknown channels from observations. 
	Various sparse recovery algorithms have been proposed, and each has its own tradeoff between computational complexity and recovery accuracy.
	The third component, which affects the recovery performance significantly, is the measurement matrix that maps the unknown sparse channels onto a compact subspace to obtain lower-dimensional linear measurements (observations). 
	Existing sparse channel estimation schemes commonly adopt random matrices \footnote{The random measurement matrix is a matrix having the elements produced by a random variable following a specified distribution. The entries of such a random measurement matrix can be independently drawn from a distribution such as the Gaussian distribution or symmetric Bernoulli distribution. The randomly chosen rows of a Fourier matrix or an orthonormal matrix can also be used to construct a measurement matrix.} as the measurement matrix, but random measurement matrices are widely known to be suboptimal\cite{choi2017compressed,ShenCompressed2016,Berger2010application, qin2018sparse,alkhateeby2015compressed}. Random measurement matrices perform random projections and cannot fully exploit underlying channel structures. Thus, they often have unsatisfactory reconstruction performance, especially when the number of measurements is insufficient \cite{Berger2010application}. While numerous advanced sparse reconstruction algorithms have been proposed, further improvement of sparse reconstructions is limited by suboptimal random measurement matrices.
	\bb{Several efforts have been made to optimize the measurement matrix \cite{qin2018sparse, ShenCompressed2016, Berger2010application, alkhateeby2015compressed, choi2017compressed}; however, it is generally challenging to optimize a matrix having hundreds to thousands of elements. An alternative to the random matrix is to design a deterministic matrix \cite{DeVore2007deterministic, Bourgain2011explicit, Xu2011deterministic} that should satisfy the restricted isometry property (RIP) to achieve accurate sparse reconstructions. However, it is non-deterministic polynomial-time hardness (NP-hard) to determine explicitly whether a matrix satisfies the RIP \cite{Berger2010application, choi2017compressed}. Some deterministic measurement matrices that approximately satisfy the RIP were designed for specific applications in an ad hoc manner \cite{obermeier2017sensing, lotfi2018a}, and do not perform well in reconstructions for different channel realizations. While traditional optimization methods encounter challenges to optimize a large-size matrix without explicit guidelines, the deep learning technique is promising to address the non-trivial optimization of measurement matrices.}

\bb{This paper aims to address the measurement matrix design issue for massive MIMO sparse channel estimations.  Based on the assumption of the standard beamspace sparse channel representation, our goal is to develop and validate a data-driven solution for designing measurement matrices that benefit the existing general sparse reconstruction algorithms. We will validate that the measurement matrices designed by data-driven solution offer superior reconstruction performance over the commonly used random matrices. }
\bb{To this end, the deep unfolding technique \cite{hershey2014deep} is used to build model-based autoencoders, and the built autoencoders are trained to acquire the data-driven measurement matrices. Then, the acquired measurement matrices are tested and evaluated by several traditional sparse reconstruction algorithms.}	
Deep unfolding is known as an attractive technique to transform a traditional model-based algorithm into a learning algorithm (or a deep learning network) by regarding a finite number of iterations as multiple-stacking layers of a deep learning network.
\bb{The examples of deep unfolding networks for sparse reconstructions include, but are not limited to, the learned iterative shrinkage-thresholding algorithm (LISTA) \cite{Gregor2010Learning, Monga2021algorithm} and the learned denoising-based approximate message passing (LDAMP) network \cite{metzler2017learned, he2018deep}.}  
\bb{When compared with other deep unfolding applications that primarily aim to develop learnable sparse reconstruction algorithms \cite{Monga2021algorithm, Gregor2010Learning, hershey2014deep, metzler2017learned, he2018deep, Magoarou2020online}, this work focuses on exploring data-driven solutions to design measurement matrix. More importantly, the acquired data-driven measurement matrices are expected to outperform the commonly used random matrices when directly applied to general conventional sparse reconstruction algorithms since they are fully optimized when given a certain dataset by training the customized autoencoders.}
	In our approach, we unfold the linear mapping as the encoder and an iterative solution of basis pursuit sparse reconstruction \cite{donoho2006compressed} as the decoder to propose a generic framework of deep unfolding basis pursuit autoencoder ($BP$-$AE$) that can learn an optimal measurement matrix via training on the given beamspace channel dataset. 
	We make crucial observations that, because a compressive sensing process can be viewed as a feedforward computation of an autoencoder, we can customize an autoencoder to mimic the compressive sensing and recovery process; more importantly, the measurement matrix, which plays an important role in computing compressive sensing and recovery, can be treated as the trainable weights throughout the customized autoencoder. These observations open up the opportunity to embed the domain knowledge of certain sparse reconstruction algorithm into deep network so that the weights (corresponding to the elements of the measurement matrix) can be optimized via the back propagation mechanism. 

\bb{A variety of model-free deep learning-based CSI acquisition approaches have been proposed and they adopt generic deep network architectures, such as the deep neural network (DNN), convolutional neural network (CNN) and recurrent neural network (RNN), and have empirically shown satisfactory performance \cite{wang2019deep, wen2018deep, Yang2019Deep, chun2019deep, chun2019deepjoint, huang2018deep, LuMIMO2019, JangDeep2019, LiuExploiting2019, Ma2020data}. Also, the reinforcement learning technique was adopted to develop sparse recovery algorithms by formulating the sparse recovery process as a sequential decision making problem \cite{Zhong2019Learning}. 
Unlike these machine learning methods, our proposed deep learning method is a hybrid and model-based data-driven approach.
The proposed method is ``hybrid data-driven" because, instead of directly using the trained autoencoders to implement sparse channel reconstructions, the autoencoders are only used to acquire data-driven measurement matrices that are then applied to conventional sparse reconstruction algorithms to accomplish the desired channel estimation task. This hybrid data-driven scheme is flexible to implement in practical channel estimations since numerous sparse reconstruction algorithms can be chosen to achieve accurate reconstructions for each individual channel realization. 
Furthermore, the proposed autoencoders are model-based since the domain knowledge of traditional sparse reconstruction algorithms are integrated to build deep unfolding autoencoders. Compared to the model-free deep learning models, which can lead to large-sized deep networks and often require a vast amount of training data, the deep unfolding autoencoders show faster convergence, have better interpretation, and are easier to train, even with a small amount of data. 
}

	The major contributions of this paper are summarized as follows:
\begin{itemize}
	\item 
	\bb{We propose a data-driven solution to address the suboptimal measurement matrix issue for sparse channel reconstructions. This is the first time a hybrid data-driven approach is developed to reveal the potential that numerous conventional channel sparse reconstruction algorithms can be significantly improved by simply changing the random measurement matrices to proper data-driven matrices.}

	\item 
	\bb{We propose a model-based autoencoder framework $BP$-$AE$ that is specially customized to optimize the measurement matrix. The $BP$-$AE$ incorporates the wisdom from the traditional basis pursuit sparse reconstruction algorithm and is customized to have the measurement matrix as the tied weights for the entire autoencoder such that the measurement matrix can be optimized via back propagations.}
	
	 \item 
	 We propose four explicit autoencoder models to acquire different measurement matrices. \bb{By comparing different models, we reveal several new techniques to improve the autoencoder performance, such as introducing pseudo residual learning units into the deep network and introducing an auxiliary nonnegativity feature into the training data.} 

	\item 
	We evaluate the acquired data-driven measurement matrices using several classical sparse reconstruction algorithms, including linear programming \cite{Candes2005decoding}, gradient projection sparse reconstruction (GPSR) \cite{figueiredo2007gradient}, and the DC-GPSR algorithms \cite{Wu2020sparse, Wu2021nonconvex}. \bb{The results confirm that the proposed data-driven measurement matrices have robust performance with different sparse reconstruction algorithms.}

\end{itemize}	

	This paper is organized into five sections.
	Section \RomanNumeralCaps{2} presents the system model of beamspace channel estimation.
	Section \RomanNumeralCaps{3} discusses the feasibility and applications of designing a real-valued measurement matrix for CSI acquisition and proposes the framework of deep unfolding basis pursuit autoencoders. 
	Then, this section proposes four explicit autoencoder models that can be trained to acquire data-driven measurement matrices for beamspace channel reconstructions. 
	Section \RomanNumeralCaps{4} presents numerical results, and Section \RomanNumeralCaps{5} concludes the paper.

%

\section{System Model}
\subsection{Sparse Beamspace Channel Estimation in Massive MIMO Systems}
	We consider a downlink massive MIMO system with a digital architecture, where the BS has $N$ antennas and each user is equipped with a single antenna.
	We consider narrowband block fading channels, and use the vector $\mathbf h_s \in \mathbb{C}^{N}$ to denote the \mbox{spatial-domain} channel between the BS and a UE \cite{gao2017reliable} 
	\begin{IEEEeqnarray*}{lCl}\label{spatial channel vector}
	\mathbf h_s = \sqrt{\frac{N}{N_p}} \sum _{l=1} ^{N_p} \beta^{(l)} \bm \alpha (\phi ^{(l)}) \IEEEyesnumber
	\end{IEEEeqnarray*}
    where $N_p$ is the number of scattering clusters; $l = 1$ is the index for the line-of-sight path; $2 \leq l \leq N_p$ is the index for non-line-of-sight paths; $\beta^{(l)}$ is the complex path gain;
    $\bm \alpha (\phi  ^{(l)})$ is the array steering vector that contains a list of complex spatial sinusoids representing the relative phase shifts of the incident far-field waveform across the array elements.
	For an $N$-element uniform linear array, the array steering vector $\bm \alpha (\phi  ^{(l)})$ can be represented by
    \begin{IEEEeqnarray*}{lCl}
    \label{steering vector}
    \bm \alpha (\phi ^{(l)}) = \frac{1}{\sqrt{N}} [1,  e^{-j2\pi \phi^{(l)}}, ... , e^{-j2\pi \phi^{(l)} (N-1)}]^{T}
    \IEEEyesnumber
    \end{IEEEeqnarray*}	 
	where $\phi ^{(l)}$ denotes the spatial direction of the $l$th path, and it is related to the physical angle $\theta^{(l)}$ by $\phi ^{(l)} = \frac{d}{\lambda} \sin{\theta ^{(l)}}$ for $-\frac{1}{2} \leq \phi ^{(l)} \leq \frac{1}{2}$ and $-\frac{\pi}{2} \leq \theta ^{(l)} \leq \frac{\pi}{2}$, where $\lambda$ is the wavelength, and $d =\frac{\lambda}{2}$ is the antenna spacing. 
	
	The spatial channel vector $\mathbf h_s$ in \eqref{spatial channel vector} can be transformed into the beamspace channel representation $\mathbf h_b$ by \cite{brady2013beamspace}
	\begin{IEEEeqnarray*}{lCl} \label{beamspace_channel_vector}
	\mathbf h_b = \mathbf{U} \mathbf h_s 
	\IEEEyesnumber
	\end{IEEEeqnarray*}    
	where $\mathbf U$ can be expressed using a set of orthogonal array steering vectors as
	\begin{IEEEeqnarray*}{lCl}
	\label{U_matrix}
	\mathbf{U}  = [\bm \alpha(\phi_1),  \bm \alpha(\phi_2),  ...  , \bm \alpha(\phi_N)]^H
	\IEEEyesnumber
	\end{IEEEeqnarray*}  
   and where $\phi_m = \frac{1}{N} (m-\frac{N+1}{2})$ for $m=1,2,...,N$ is the spatial direction predefined by the array having half-wavelength spaced antennas. From \eqref{U_matrix}, we can see that the matrix $\mathbf U$ is simply a DFT matrix having the size of $N \times N$.
	The beamspace channel $\mathbf h_b$ is sparse.
	In practice, due to the limited resolution of virtual angular bins, mismatches can happen between the real angle of departures (AoDs) and the predefined virtual angular bins. 
   This phenomenon is known as power leakage and it causes the beamspace channels to be imperfectly sparse. 
   However, power leakage is negligible for a massive MIMO system having hundreds of antenna elements.
   For this situation, we say that the beamspace channel is approximately sparse or compressible, since the energy (square sum) of a few large-valued elements can capture the most of the channel energy $\norm{\mathbf{h}_b}_2^2$ \cite{heath2016an}.
  In this paper, we neglect the power leakage and assume the beamspace channels are perfectly sparse by treating the small-valued elements as zeros.
	For pilot-aided downlink channel estimation, the BS transmits the known pilots represented by matrix $\mathbf{P}$ to users. 
	The UE receives the pilot sequence 
    \begin{IEEEeqnarray*}{lCl}
    \label{received_pilots}
    \mathbf{r} = \mathbf{P}\mathbf{h}_s + \mathbf{w} 
    \IEEEyesnumber
    \end{IEEEeqnarray*}
	where $\mathbf{r} \in \mathbb{C}^{M}$ is the received pilot sequence, $\mathbf P \in \mathbb{C}^{M \times N}$ contains the pilot sequences for $N$ antennas transmitted over $M$ time slots, $\mathbf{h}_s \in \mathbb{C}^{N}$ is the spatial-domain channel vector, $\mathbf{w} $ is the received noise vector and $\mathbf{w} \sim \mathcal{CN}(0, \sigma_n^2 \mathbf I)$.
	Conventional linear reconstruction methods, such as linear minimum mean square error estimation or least squares (LS), require $M \geq N$ for robust channel estimations, where $M$ is the length of pilot sequences and $N$ is the number of BS antennas.
	This requirement on pilot length will cause prohibitive spectral occupancy and high computational complexity for massive MIMO systems having a large number of antennas.
	For this reason, it is attractive to develop the compressive-sensing-aided beamspace channel estimation schemes \cite{choi2017compressed}.
	 According to the relationship between the spatial-domain channel $\mathbf{h}_s$ and the beamspace channel $\mathbf h_b$ in \eqref{beamspace_channel_vector}, we replace the channel vector $\mathbf{h}_s$ by $\mathbf h_s = \mathbf U^{H} \mathbf h_b$ and rewrite the received pilot symbols in \eqref{received_pilots} as
    \begin{IEEEeqnarray*}{lCl}
    \label{received_pilots_beamspace}
    \mathbf{r} =  \mathbf{P} \mathbf U^{H} \mathbf h_b + \mathbf{w}.
    \IEEEyesnumber
    \end{IEEEeqnarray*}
    Next, denoting $\mathbf{\Phi} =\mathbf{P} \mathbf U^{H}$ allows \eqref{received_pilots_beamspace} to be expressed as 
     \begin{IEEEeqnarray*}{lCl}
     \label{CS_estimation}
    \mathbf{r} =\mathbf{\Phi} \mathbf h_b + \mathbf{w}.
     \IEEEyesnumber  
    \end{IEEEeqnarray*}
     According to the compressive sensing theory, the sparse beamspace channel vector $\mathbf h_b$ can be estimated accurately from the measurements $\mathbf r$ for $M \ll N$ \cite{choi2017compressed}, thus the overheads of downlink pilots and CSI feedback can be largely reduced.	
	
	By stacking the real part and the imaginary part of received pilots, we can express \eqref{CS_estimation} as
	\begin{IEEEeqnarray*}{lCl}
	\label{real_and_imag}
	\begin{bmatrix}
	\Re(\mathbf{r}) \\ \Im(\mathbf{r})
	\end{bmatrix}
	= \begin{bmatrix}
	\Re(\mathbf \Phi) & -\Im(\mathbf \Phi) \\
	\Im(\mathbf \Phi) & \Re(\mathbf \Phi) 
	\end{bmatrix}
	\begin{bmatrix}
	\Re(\mathbf h_b) \\ \Im(\mathbf h_b)
	\end{bmatrix}
	+ 	
	\begin{bmatrix}
	\Re(\mathbf{w}) \\ \Im(\mathbf{w})
	\end{bmatrix}
	\IEEEyesnumber
	\end{IEEEeqnarray*}
	where $\Re(\cdot)$ and $\Im(\cdot)$ denote the real part and imaginary part of a complex matrix or a vector. 
	In this paper, we adopt the design criterion to force the measurement matrix to take real values $\mathbf{\Phi} \in \mathbb R^{M \times N}$, i.e., $\Im(\mathbf \Phi) = \mathbf 0$, $\mathbf \Phi = \Re(\mathbf \Phi)$.
    Thus, \mbox{eq. \eqref{real_and_imag}} can be expressed as
	\begin{IEEEeqnarray*}{lCl}
	\label{real_sepa_imag}
	\begin{bmatrix}
	\Re(\mathbf{r}) \\ \Im(\mathbf{r})
	\end{bmatrix}
	= \begin{bmatrix}
	\mathbf \Phi & \mathbf 0 \\
	\mathbf 0 & \mathbf \Phi 
	\end{bmatrix}
	\begin{bmatrix}
	\Re(\mathbf h_b) \\ \Im(\mathbf h_b)
	\end{bmatrix}
	+ 	
	\begin{bmatrix}
	\Re(\mathbf{w}) \\ \Im(\mathbf{w})
	\end{bmatrix}
	\IEEEyesnumber.
	\end{IEEEeqnarray*} 
	Equation \eqref{real_sepa_imag} can be equivalently expressed by
     \begin{IEEEeqnarray*}{lCl}
	\label{CS_separate}
	[\Re(\mathbf{r}),\Im(\mathbf{r})] &= \mathbf \Phi  [\Re(\mathbf{h}_b), \Im(\mathbf{h}_b)] + [\Re(\mathbf{w}),\Im(\mathbf{w})]. 
	\IEEEyesnumber
	\end{IEEEeqnarray*}
	We further write \eqref{CS_separate} as
	\begin{IEEEeqnarray*}{lCl}
	\label{CS_unified}
	\mathbf{R} =  \mathbf{\Phi} \mathbf{H} + \mathbf{W}
	\IEEEyesnumber
	\end{IEEEeqnarray*}
    where $\mathbf{R} = [\Re(\mathbf{r}), \Im(\mathbf{r})] \in \mathbb{R}^{M \times 2}$ contains the measurements;  $\mathbf{\Phi} \in \mathbb{R} ^ {M \times N}$ is the measurement matrix that we aim to design;
     $\mathbf{H}=[\Re(\mathbf h_b), \Im(\mathbf h_b)] \in \mathbb{R}^{N \times 2}$ represents a sparse beamspace channel to be estimated, and it is a column-wise stacking of the real part and imaginary part of a complex beamspace channel vector;
    $\mathbf{W}=[\Re(\mathbf w), \Im(\mathbf w)] \in \mathbb{R}^{M \times 2}$ contains noise vectors. 
    In the remainder of this paper, we refer the real-form matrix $\mathbf{H}$ as a sample of sparse beamspace channels and it corresponds to a complex-valued beamspace channel vector $\mathbf{h}_b$.

\section{Learning Measurement Matrices}
	For the application of either pilot design or compressed CSI feedback, it is essential to design a suitable measurement matrix $\mathbf \Phi$ to reconstruct the sparse beamspace channels accurately, since the choice of the measurement matrix highly influences channel reconstruction performance.
	In this section, we present a data-driven solution to design the measurement matrix. 
	We first discuss the feasibility of real-valued measurement matrices and the applications in massive MIMO channel estimations. Then, we propose the model-based autoencoder framework $BP\textit{-}AE$, which is designed by unfolding the process of linear compressive sensing and nonlinear basis pursuit sparse reconstruction. 
Under this framework, we propose two basic autoencoder models and two extension models that are set to have the measurement matrix as the tied weights to be trained to acquire different data-driven measurement matrices. 

	\begin{figure}[!t]
	\centering
	\includegraphics[scale=0.3]{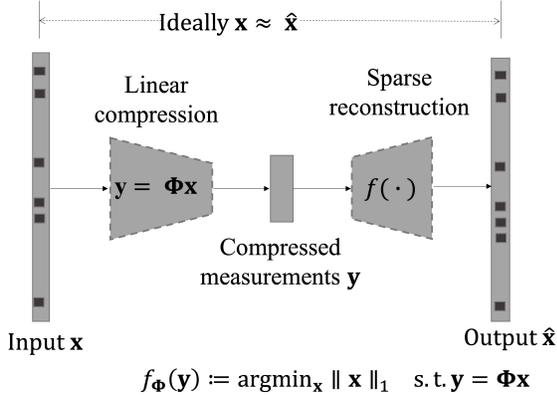}
	\caption{Framework of deep unfolding $BP\textit{-}AE$}
	\label{cs_autoencoder}
	\end{figure}	

\subsection{Feasibility of Real-Valued Measurement Matrix $\mathbf \Phi$ and Applications in Channel Estimation}
   
    \bb{In this work, we propose to design real-valued measurement matrices. The reconstruction performance of the real-valued matrices, such as the Gaussian matrix or Bernoulli matrix, were tested and found to be similar to the reconstruction performance of the complex-valued random Fourier matrix. Moreover, when adopting real-valued measurement matrices by forcing the imaginary part of $\mathbf \Phi$ to be zeros, as shown in \eqref{real_and_imag} and \eqref{real_sepa_imag}, the computational complexity reduces to a half of the original complex-valued matrix-vector multiplication. Thus, we conclude that the real-valued measurement matrix design can reduce the computational complexity without degrading the reconstruction performance.} 
	We consider two application scenarios to apply the real-valued measurement matrix $\mathbf \Phi$ to downlink CSI acquisitions in FDD massive MIMO systems.
	The first application is to design downlink pilots. 
	 Based on \eqref{received_pilots_beamspace} and \eqref{CS_estimation}, the pilot matrix $\mathbf P$ and the measurement matrix $\mathbf \Phi$ can be related by $\mathbf P = \mathbf \Phi \mathbf{U}$, where the matrix $\mathbf U$ is a DFT matrix.
	 Thus, designing a complex-valued \footnote{Note that the pilot matrix $\mathbf P$ is complex valued because $\mathbf U$ is a complex-valued unitary matrix.} pilot matrix $\mathbf P$ is equivalent to designing a real-valued measurement matrix $\mathbf \Phi$. 
	A practical pilot matrix design needs to consider a power constraint, and ideally each column of $\mathbf P$ is expected to have a unit $\ell_2$-norm to satisfy the power constraint on the pilot sequences transmitted by each BS antenna.
	 This power constraint on the pilot matrix can be equivalently accomplished by normalizing the columns of the measurement matrix $\mathbf \Phi$.
	 Unless specified otherwise, our acquired data-driven measurement matrices will be normalized column-wise before being used for beamspace channel reconstructions.
	The second application of the measurement matrix $\mathbf \Phi$ is CSI feedback reduction.
	If ideal CSI is already known at the UE via perfect channel estimations, then the UE can linearly compress the beamspace channel $\mathbf H$ by simply multiplexing the measurement matrix $\mathbf{\Phi}$. Instead of transmitting the high-dimensional beamspace channel $\mathbf H$, the UE sends the compressed measurements $\mathbf R = \mathbf{\Phi} \mathbf H $ back to the BS.
	At the end of the process, the beamspace channels can be reconstructed at the BS. This process reduces the CSI feedback overhead significantly.

\subsection{Framework of Deep Unfolding \textit{BP\textit{-}AE}}	
	For a beamspace channel dataset $\{\mathbf X \}$ containing $n$ beamspace channel samples, we represent this dataset by
	\begin{equation}
	\begin{aligned}
	 \{[\mathbf H_1,\mathbf H_2,...,\mathbf H_n] \} 
	=&  \{[\Re(\mathbf h_{b,1}), \Im(\mathbf h_{b,1}), \Re(\mathbf h_{b,2}), \Im(\mathbf h_{b,2}),  \\
	& ..., \Re(\mathbf h_{b,n}), \Im(\mathbf h_{b,n})]\}.
	\end{aligned}	
	\end{equation}
	Without loss of generalization, we refer to the vector $\mathbf x \in \mathbb{R}^{N}$ as an arbitrary column in the dataset $\{\mathbf X\}$, and thus the vector $\mathbf x$ represents either $\Re(\mathbf h_{b,i})$ or $\Im(\mathbf h_{b,i})$ for $1 \le i \le n$. 
	Then, we can express the sparse reconstruction problem for noisy scenarios using the underdetermined equation 
	\begin{equation}
	\mathbf y = \mathbf \Phi \mathbf x + \mathbf n
	\end{equation}
	where $\mathbf y \in \mathbb{R}^{M}$ is the measurement vector, $\mathbf x \in \mathbb{R}^{N}$ is a sparse vector and $\mathbf n \in \mathbb{R}^{M}$ is the noise vector.
	We will design an autoencoder that can learn the latent representation (code) $\mathbf y$ for input data $\mathbf x$, and can subsequently output the reconstruction $\hat{\mathbf x}$ such that $\hat{\mathbf x} \approx \mathbf x$.
	An autoencoder consists of two main parts: an encoder that maps the input into the code, i.e., $\mathbf y = g (\mathbf x)$, and a decoder that maps the code to the reconstruction of the original input, i.e., $\hat{\mathbf x} = f(\mathbf y)$.
	 Here, functions $g(\cdot)$ and $f(\cdot)$ represent the overall nonlinear transformations of the encoder and the decoder respectively.
	 Based on an observation that the compressive sensing and recovery process can be regarded as a realization of a feedforward computation of an autoencoder, we propose an autoencoder framework called deep unfolding $BP\textit{-}AE$, which mimics the process of linear compression and basis pursuit sparse reconstruction. 
	As shown in \mbox{Fig. \ref{cs_autoencoder}}, the deep unfolding $BP\textit{-}AE$ consists of a linear encoder and a nonlinear decoder, which are represented by
	\begin{equation}
	\label{linear_encode}
	g_{\{\mathbf \Phi \}}(\mathbf x) := \mathbf \Phi \mathbf x
	\end{equation}	
	\begin{equation}
	\label{bp}
	f_{\{\mathbf \Phi \}}(\mathbf y) := \mathop{\text{argmin}}\limits_{\mathbf x} {\norm{\mathbf x}_1} \quad \text{s.t.} \quad \mathbf y = \mathbf \Phi \mathbf x
	\end{equation}	
	where $\norm{\cdot}_1$ represents the $\ell_1$-norm, which is defined as the sum of absolute values of all elements in a vector. The encoder $g_{\{\mathbf \Phi \}} (\mathbf x)$ in \eqref{linear_encode} performs a linear dimension reduction; the decoder $f_{\{\mathbf \Phi \}}(\mathbf y)$ mimics solving an \mbox{$\ell_1$-minimization} sparse reconstruction in \eqref{bp}, which is also known as the basis pursuit in signal processing. 
	\bb{We choose to unfold the basis pursuit algorithm because the linear constraint in \eqref{bp} can strictly force the measurement matrix to perform a proper dimensionality-reduction mapping and retain more useful information of the sparse vectors.}
	Another important observation is that we can treat the measurement matrix as the weights used within the deep unfolding $BP\textit{-}AE$.
	Therefore, we can parameterize the autoencoder with a trainable measurement matrix $\{\mathbf \Phi\}$ as the tied weights of both the encoder and the decoder, so that the measurement matrix can be optimized during model training with a given dataset by back propagation algorithms.
	The loss function of the deep unfolding $BP\textit{-}AE$ is defined as the mean squared error (MSE) between input samples and output reconstructed vectors, 
	\begin{equation}
	\label{MSE}
	\mathcal L(\mathbf x, \hat{\mathbf x}) = \frac{1}{n} \sum _{i=1}^{n} \norm{\mathbf{x}_i - \hat{\mathbf x_i}}_2 ^2 	
	\end{equation}
where $n$ is the number of training samples in dataset $\{\mathbf X\}$, $\mathbf{x}_i$ represents the $i$th input sample and $\hat{\mathbf x_i}$ represents the $i$th reconstructed vector. From the viewpoint of feedforward computation, the autoencoder performs a linear dimension reduction jointly with a basis pursuit sparse reconstruction, thus it can be interpreted as a compressive sensing and reconstruction process.
	From the viewpoint of backward propagation, the autoencoder back propagates and minimizes the reconstruction error in \eqref{MSE} by optimizing the measurement matrix, thus it can be interpreted as a stochastic optimizer for the measurement matrix. 	
	The optimizer can adopt any standard back propagation algorithm in deep learning such as the stochastic gradient descent (SGD) algorithm \cite{qin2019deep}.
	After training, we can extract the weights $\mathbf \Phi \in \mathbb{R}^{M \times N}$ from the trained autoencoder and save it as the acquired data-driven measurement matrix.

\subsection{Specific Structures of Deep Unfolding \textit{BP\textit{-}AE}}

	\begin{figure}[!t]
	\centering
	\includegraphics[scale=0.25]{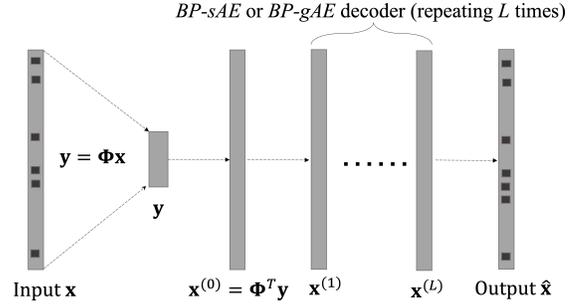}
	\caption{Diagram showing the structure of deep $BP\textit{-}AE$}
	\label{unfolding}
	\end{figure}
	
	The overall network structure of deep unfolding $BP\textit{-}AE$ is shown in \mbox{Fig. \ref{unfolding}}.
	The encoder is simply a linear fully-connected layer without an activation function as shown in \eqref{linear_encode}.
	The measurements $\mathbf y$ are the input of the decoder \footnote{\bb{We suggest to use noiseless measurements in the proposed autoencoders, since we observed that the reconstruction performance negligibly changed after adding an additive Gaussian noise layer to the encoder, but the training time increased significantly.}}. 
	The first-layer decoder is set as $\mathbf x^{(0)} = \mathbf \Phi^T \mathbf y$.
	For the decoder layers from $\mathbf x^{(1)}$ to $\mathbf x^{(L)}$, the network is designed by stacking the unfolding \footnote{The method of using an iterative solution to construct a deep neural network is called deep unfolding \cite{hershey2014deep}; this method regards each step of iterations as a single layer of a neural network and is used to transform a traditional iterative algorithm into a stack layers for a neural network. Unlike traditional iterative algorithms having predefined stopping conditions and manually-tuned parameters, the unfolding neural networks have a fixed number of iterative layers and the trainable parameters are learned from the training data.} iterations of sparse reconstruction, which can be represented by $\mathbf x^{(t+1)} = f_{\mathbf \Phi}(\mathbf x^{(t)}, \mathbf y)$ for \mbox{$0 \le t \le L-1$}.
	 We derive an iterative solution by solving the basis pursuit sparse reconstruction, then we can specify the decoder structures by unfolding the derived iterations.


	To solve the basis pursuit sparse reconstruction problem in \eqref{bp}, we adopt the projected subgradient descent method \cite{boyd2003subgradient}, and the update is given by 
	\begin{equation}
	\label{projected_subgradient_method}
	\mathbf x^{(t+1)} = \textit{Proj}\big(\mathbf x^{(t)} - \alpha _t \cdot \text{sign} (\mathbf x^{(t)}) \big)	
	\end{equation}
	where $t > 0$ indicates the $t$th update, $\alpha _t$ is the step size, $\text{sign}(\cdot)$ represents the sign function that is the subgradient of the $\ell_1$-norm term $\norm{\cdot}_1$ and $\textit{Proj}(\cdot)$ represents the projection operation onto the convex set \mbox{$\{ \mathbf x^\prime: \mathbf \Phi \mathbf x^\prime = \mathbf y \}$}.	
	The projection of a vector $\mathbf z$ onto the set $\{ \mathbf{x^\prime}: \mathbf \Phi \mathbf x^\prime = \mathbf y\}$ has the following \mbox{closed-form} solution
	\begin{equation}	 
	\label{projection_x}
	\textit{Proj}(\mathbf z) = \mathbf z + \mathbf \Phi ^\dagger (\mathbf y - \mathbf \Phi \mathbf z) 	
	\end{equation}
where $\mathbf \Phi^\dagger  = \mathbf \Phi^T(\mathbf \Phi \mathbf \Phi^T)^{-1}$ is the Moore-Penrose pseudo-inverse of $\mathbf \Phi$. 
	From \eqref{projected_subgradient_method} and \eqref{projection_x}, we substitute the vector $\mathbf{z} = \mathbf{x}^{(t)} - \alpha _t \cdot \text{sign}(\mathbf {x}^{(t)})$ into the projection operation \eqref{projection_x} and obtain the $t$th-step update as
	\begin{equation}
	\label{projected_subgradient_update}
	\begin{aligned}
	\mathbf x ^{(t+1)} 
	&=& \mathbf x^{(t)} + \mathbf \Phi ^\dagger( \mathbf y -	\mathbf \Phi \mathbf x ^{(t)}) 
	 - \alpha _t (\mathbf{I}-\mathbf {\Phi}^\dagger \mathbf {\Phi}) \cdot\text{sign}(\mathbf x^ {(t)}).
	 \end{aligned}
	\end{equation}
	Equation \eqref{projected_subgradient_update} is an iterative solution of the basis pursuit sparse reconstruction in \eqref{bp}.	
	The step size parameter can be set as $\alpha_t=\alpha/t$ according to the diminishing step size rule \cite{boyd2003subgradient}.
	Given a proper starting point $\mathbf x^{(1)}$ and a stopping condition, the update \eqref{projected_subgradient_update} can be used to obtain the reconstruction $\hat{\mathbf x}$.
	 To reduce the intensive computations required to calculate the pseudo-inverse $\mathbf \Phi^\dagger$, we will simplify the pseudo-inverse computation by replacing $\mathbf \Phi^\dagger$ with the transpose $\mathbf \Phi ^T$.
	Thus, the decoder update $\mathbf x^{(t+1)} = f_{\mathbf \Phi}(\mathbf x^{(t)}, \mathbf y)$ for $1 \leq t \leq L-1$ can be represented by
	\begin{equation}
	\label{unfold_iteration}
	\begin{aligned}
	 \mathbf x^{(t+1)} 
	 = & \mathbf x^{(t)} + \mathbf \Phi ^T \mathbf y - \mathbf \Phi^{T} 	\mathbf \Phi \mathbf x ^{(t)} 
	 \\ &  -(\alpha/t) (\mathbf{I}-\bm{\Phi}^{T} \bm{\Phi}) \cdot \text{sign}(\mathbf x^ {(t)}).
	  \end{aligned}
	\end{equation}
	We design the decoder structure $\mathbf x^{(t+1)} = f_{\mathbf \Phi}(\mathbf x^{(t)}, \mathbf y)$ for \mbox{$0 \le t \le L-1$} by unfolding the iterations of \eqref{unfold_iteration}.
	 By specifying the measurements $\mathbf y$ in different ways, we construct two explicit decoder structures named the $BP\textit{-}sAE$ decoder and the $BP\textit{-}gAE$ decoder, as explained below.

	\begin{figure}[!t]
	\centering
	\includegraphics[scale=0.3]{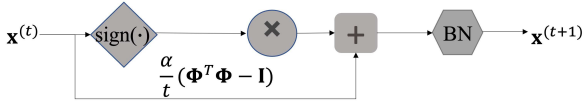}
	\captionsetup{justification=centering}
	\caption{Computation graph of a $BP\textit{-}sAE$ decoder; the module BN is batch normalization}
	\label{l1sae}
	\end{figure}
	
	\begin{figure}[!t]
	\centering
	\includegraphics[scale=0.3]{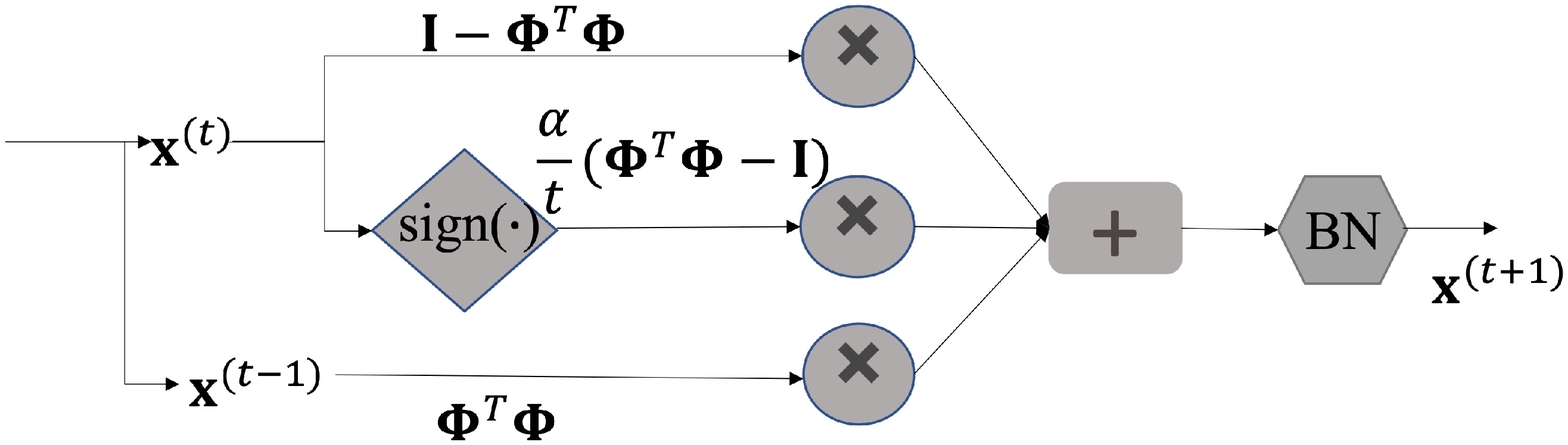}
	\captionsetup{justification=centering}
	\caption{Computation graph of a $BP\textit{-}gAE$ decoder; the module BN is batch normalization}
	\label{l1gae}
	\end{figure}

\textit{$BP\textit{-}sAE$ Decoder:}
	We specify $\mathbf y$ as the measurements only in current layer, i.e., $\mathbf y=\mathbf \Phi \mathbf x^{(t)}$.
	By substituting $\mathbf y=\mathbf \Phi \mathbf x^{(t)}$ into the decoder update \eqref{unfold_iteration}, we obtain a simplified update
	\begin{equation}
	\label{l1sae_decoder}
	\mathbf x^{(t+1)} = \mathbf x^{(t)} -(\alpha/t)(\mathbf{I}-\bm{\Phi}^{T} \bm{\Phi}) \cdot \text{sign}(\mathbf x^ {(t)}).
	\end{equation}
	Equation \eqref{l1sae_decoder} is the $t$th-layer computation for the $BP\textit{-}sAE$ decoder, and its computation graph\footnote{In deep learning, the computation graph is used to represent a math function in the language of graph theory for visualizing the computation flow.} is shown in Fig. \ref{l1sae}.

\emph{$BP\textit{-}gAE$ Decoder:}
	We specify $\mathbf y$ as the measurements obtained in previous layer, i.e., $\mathbf y=\mathbf \Phi \mathbf x^{(t-1)}$.
	By substituting $\mathbf y=\mathbf \Phi \mathbf x^{(t-1)}$ in the decoder update \eqref{unfold_iteration}, we obtain
	\begin{equation}
	\label{l1gae_decoder}
	\begin{aligned}
	\mathbf x^{(t+1)} = & \mathbf x^{(t)} + \mathbf \Phi ^{T} \mathbf \Phi \mathbf x^{(t-1)} - \mathbf \Phi^{T} 	\mathbf \Phi \mathbf x^{(t)} 
	 \\ & -(\alpha/t) (\mathbf{I}-\bm{\Phi}^{T} \bm{\Phi}) \cdot \text{sign}(\mathbf x^ {(t)}).
	 \end{aligned}
	\end{equation}
	 Equation \eqref{l1gae_decoder} is the $t$th-layer computation for the $BP\textit{-}gAE$ decoder, and its computation graph is shown in Fig. \ref{l1gae}.
	 \bb{The advantage of the $BP\textit{-}gAE$ decoder is that we introduce a shortcut from the previous output $\mathbf x^{(t-1)}$ by treating the measurements $\mathbf y$ as $\mathbf \Phi \mathbf x^{(t-1)}$. We refer to this as pseudo residual learning and it can improve the autoencoder performance since the introduced special unit $\mathbf \Phi ^{T} \mathbf \Phi \mathbf x^{(t-1)} - \mathbf \Phi^{T} \mathbf \Phi \mathbf x^{(t)}$ can be understood as a special residual learning unit \cite{he2016deep}.}

	In the $BP\textit{-}AE$ framework in Fig. \ref{unfolding}, the decoder computations from $\mathbf x^{(1)}$ to $\mathbf x^{(L)}$  are specified by repeating the computations of the $BP\textit{-}sAE$ decoder in Fig. \ref{l1sae} and the $BP\textit{-}gAE$ decoder in Fig. \ref{l1gae} for $L$ times, leading to two autoencoder models named as the $BP\textit{-}sAE$ and $BP\textit{-}gAE$.
	The output layer of the autoencoder is represented by $\hat{\mathbf x} = \mathbf x^{(L)}$.
	The loss function is the MSE between input data $\{\mathbf X\}$ and output reconstructions $\{\hat{\mathbf X}\}$, as shown in \eqref{MSE}.
	The trainable variables are $\{\mathbf \Phi, \alpha\}$, where $\mathbf \Phi$ is configured to be the tied weights of autoencoder and it is also the measurement matrix that we aim to optimize by training the autoencoder.

\subsection{Extension Models for Learning the Measurement Matrices Having Double Columns}

	\begin{figure}[!tb]
	\centering
	\includegraphics[scale=0.33]{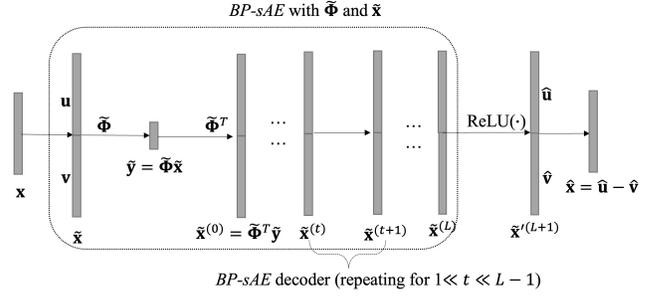}
	\captionsetup{justification=centering}
	\caption{Diagram that shows the $BP\textit{-}sAEcat$ extension model}
	\label{l1sae_cat}
	\end{figure}
	
	\begin{figure}[!tb]
	\centering
	\includegraphics[scale=0.31]{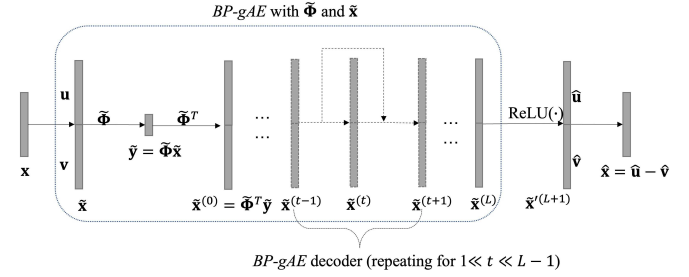}
	\captionsetup{justification=centering}
	\caption{Diagram that shows the $BP\textit{-}gAEcat$ extension model}
	\label{l1gae_cat}
	\end{figure}
	
	\bb{In the previous subsections, we have proposed the $BP\textit{-}sAE$ or $BP\textit{-}gAE$ autoencoders to acquire data-driven measurement matrices having the size of $M \times N$, where $M$ is the compressed dimension and $N$ is the dimension of sparse vectors.
	To further improve the measurement matrix performance, we introduce an artificial feature, i.e., non-negativity, to the dataset and propose two extension models that are referred to as the $BP\textit{-}sAEcat$ and $BP\textit{-}gAEcat$.
	Distinct from the $BP\textit{-}sAE$ and $BP\textit{-}gAE$, the $BP\textit{-}sAEcat$ and $BP\textit{-}gAEcat$ produce measurement matrices having the size of $M \times 2N$.}
	
	We first perform variable splitting at the input of encoder. 
	The channel vector $\mathbf x$ is represented by the difference between its positive part and negative part, i.e., $\mathbf x = \mathbf u - \mathbf v $, where
	\begin{equation}
	\label{sepa_pos_neg}
	\mathbf u =  \mathbf x_{+}, \quad
	\mathbf v =  (-\mathbf x)_{+}
	\end{equation}
	where $\mathbf x_+$ denotes retaining the positive components of vector $\mathbf x$ and setting the other elements as zeros, i.e., $(\mathbf x_+)_i = \text{max}\{\mathbf x_i, 0\}$ for $i = 1, \cdots, {N}$; $(-\mathbf x)_+$ denotes retaining the positive components of the vector $-\mathbf x$ and setting the other elements be zeros, i.e., $((-\mathbf x)_+)_i = \text{max}\{-\mathbf x_i, 0\}$ for $i = 1, \cdots, {N}$.
	Then, the problem \eqref{bp} can be rewritten as
	\begin{equation}
	\label{sepa_LP}
	\begin{aligned}
	\mathop{\min} \limits_{\mathbf u, \mathbf v} \quad & \norm{(\mathbf u - \mathbf v )}_1 \qquad  \\
	\text{s.t.} \quad & \mathbf{\Phi} (\mathbf u - \mathbf v  )  = \mathbf{y}, 
	\mathbf u \ge \mathbf{0}, \mathbf v \ge \mathbf{0}.
	\end{aligned}
	\end{equation}
	We use the double-sized channel vector $\tilde{\mathbf x}$ to denote the concatenation of $\mathbf u$ and $\mathbf v$, i.e., $\tilde{\mathbf x} = [\mathbf u^T, \mathbf v^T]^T$, and rewrite the problem \eqref{sepa_LP} as
	 \begin{equation}
	\label{standard_LP}
	\mathop{\min}\limits_{\tilde{\mathbf x}} \quad \norm{ \tilde{\mathbf x}}_1 \qquad \text{s.t.} \quad [\mathbf{\Phi}, -\mathbf{\Phi}] \tilde{\mathbf x}= \mathbf{y}, \tilde{\mathbf x}\ge \mathbf{0}.
	\end{equation}
	We relax the matrix $[\mathbf{\Phi}, -\mathbf{\Phi}]$ for $\mathbf \Phi \in \mathbb R^{M \times N}$ to the double-column sized matrix $\tilde{\mathbf \Phi} \in \mathbb R^{M \times 2N}$, and the problem \eqref{standard_LP} can be expressed as
	\begin{equation}
	\label{norm1_minimization_cat}
	\mathop {\min } \limits_{\tilde{\mathbf x}} \quad \norm{\tilde{\mathbf x}}_1 \qquad  \text{s.t.} \quad \tilde{\mathbf \Phi} \tilde{\mathbf x} = \bb{\tilde{\mathbf y}},  \tilde{\mathbf x}\ge \mathbf{0}.
	\end{equation}
	Equations \eqref{norm1_minimization_cat} and \eqref{bp} express the same sparse reconstruction problem, but they exhibit different formulations.
	First, the variables $\tilde{\mathbf x}$ and $\tilde{\mathbf \Phi}$ in problem \eqref{norm1_minimization_cat} are double-sized when compared with the variables $\mathbf x$ and $\mathbf \Phi$ in problem \eqref{bp}.
	Second, we impose an auxiliary nonnegativity constraint $\tilde{\mathbf x} \ge \mathbf{0}$ on problem \eqref{norm1_minimization_cat}.
	
	We apply the autoencoders $BP\textit{-}sAE$ and $BP\textit{-}gAE$ to the intermediate variables $\tilde{\mathbf x}$.
	At the output layer, we first apply the ReLU \footnote{\text{ReLU} is a type of activation function defined as $\text{ReLU}(x) = \text{max}(0, x)$ when $x$ is a scalar. When the input is a vector, $\text{ReLU}(\cdot)$ is applied in an element-wise manner.} activation function to $\tilde{\mathbf x}^{(L)}$ and obtain
	\begin{equation}
	\tilde{\mathbf x}^{\prime(L+1)}= \text{ReLU}\left(\tilde{\mathbf x}^{(L)}\right). 
	\end{equation}
	Then, we treat the first half of $\tilde{\mathbf x}^{\prime(L+1)}$ as the reconstruction of the positive part of $\mathbf x$, which can be expressed as $\hat{\mathbf u}= \hat{\mathbf x}_+$. Next, we treat the second half of $\tilde{\mathbf x}^{\prime(L+1)}$ as the reconstruction of the negative part of $\mathbf x$, which can be expressed as $\hat{\mathbf v} = (-\hat{\mathbf x})_+$.	
	 Thus, we obtain the final reconstruction as 
	\begin{equation}
	\label{output_relu_layer}
	\hat{\mathbf x} = \hat{\mathbf u} - \hat{\mathbf v}
	\end{equation}	
	where $\hat{\mathbf u}$ is the first-half slice of vector $\tilde{\mathbf x}^{\prime(L+1)}$, and $\hat{\mathbf v}$ is the second-half slice of vector $\tilde{\mathbf x}^{\prime(L+1)}$. \mbox{Fig. \ref{l1sae_cat}} and \mbox{Fig. \ref{l1gae_cat}} display the structures of the $BP\textit{-}sAEcat$ and $BP\textit{-}gAEcat$ extension autoencoder models respectively.

\bb{Now, we briefly analyze the computational complexity and the implementations of the proposed autoencoders. Table \ref{complexity} shows the forward-pass computational complexity of the encoders and decoders. To implement the autoencoders, we set the measurement matrix $\mathbf \Phi$ (or $\tilde{\mathbf \Phi}$) as the tied weights of the autoencoders. Thus, the number of trainable parameters is $MN$ for the $BP$-$sAE$ and $BP$-$gAE$ autoencoders and is $2MN$ for the $BP$-$sAEcat$ and $BP$-$gAEcat$ autoencoders. We can see the proposed autoencoders have a realistic computational complexity, and a practical implementation is feasible due to the reasonable amount of trainable parameters.}
Compared with the data-driven measurement matrix $\mathbf \Phi \in \mathbb{R}^{M \times N}$ acquired by $BP\textit{-}sAE$ and $BP\textit{-}gAE$, the data-driven measurement matrix $\tilde{\mathbf \Phi} \in \mathbb{R}^{M \times 2N}$ acquired by the extension models $BP\textit{-}sAEcat$ and $BP\textit{-}gAEcat$ have double-sized columns.
	 To use the measurement matrix $\tilde{\mathbf \Phi}$ in sparse reconstruction algorithms, we only need to split and concatenate the input variables as shown in \eqref{sepa_pos_neg} and then slice and subtract the output as shown in \eqref{output_relu_layer}. 

\begin{table}[t!]
  \begin{center}
  \captionsetup{justification=centering}
    \caption{Forward-Pass Computational Complexity of the Encoders and Decoders}
    \label{complexity}
    \begin{tabular}{c|c|c|c|c} 
    \hline
     & $BP\textit{-}sAE$ &  $BP\textit{-}gAE$ & $BP\textit{-}sAEcat$  & $BP\textit{-}gAEcat$ \\
    \hline
      Encoder & $O(MN)$ & $O(MN)$ & $O(2MN)$ & $O(2MN)$ \\
    \hline
      Decoder & $O(L M N^2)$ & $O(L M N^2)$ & $O(4 L M N^2)$ & $O(4 L M N^2)$ \\
     \hline
    \end{tabular}
  \end{center}
\end{table}

\vspace{-5mm}
\section{Numerical Results}

	We have proposed four deep unfolding $BP\textit{-}AE$ autoencoder models to acquire data-driven measurement matrices.
	In this section, we will evaluate the reconstruction performance of the acquired data-driven measurement matrices and compare the results with several common random matrices.
	To this end, we adopt three sparse reconstruction algorithms including linear programming recovery \cite{Candes2005decoding}, GPSR \cite{figueiredo2007gradient}, and the recently proposed DC-GPSR \cite{Wu2020sparse} to solve sparse reconstructions. 
\bb{Finally, the sparse reconstruction performance is compared with existing deep-learning based sparse reconstruction methods.}
\bb{Moreover, by comparing the different data-driven measurement matrices acquired by the proposed models, several implementation insights are discussed and the complexity-accuracy tradeoff is analyzed.}

\subsection{Experiment Setup and Autoencoder Training}

	\begin{figure*}[!tb]
	\centering
	\includegraphics[scale=0.45]{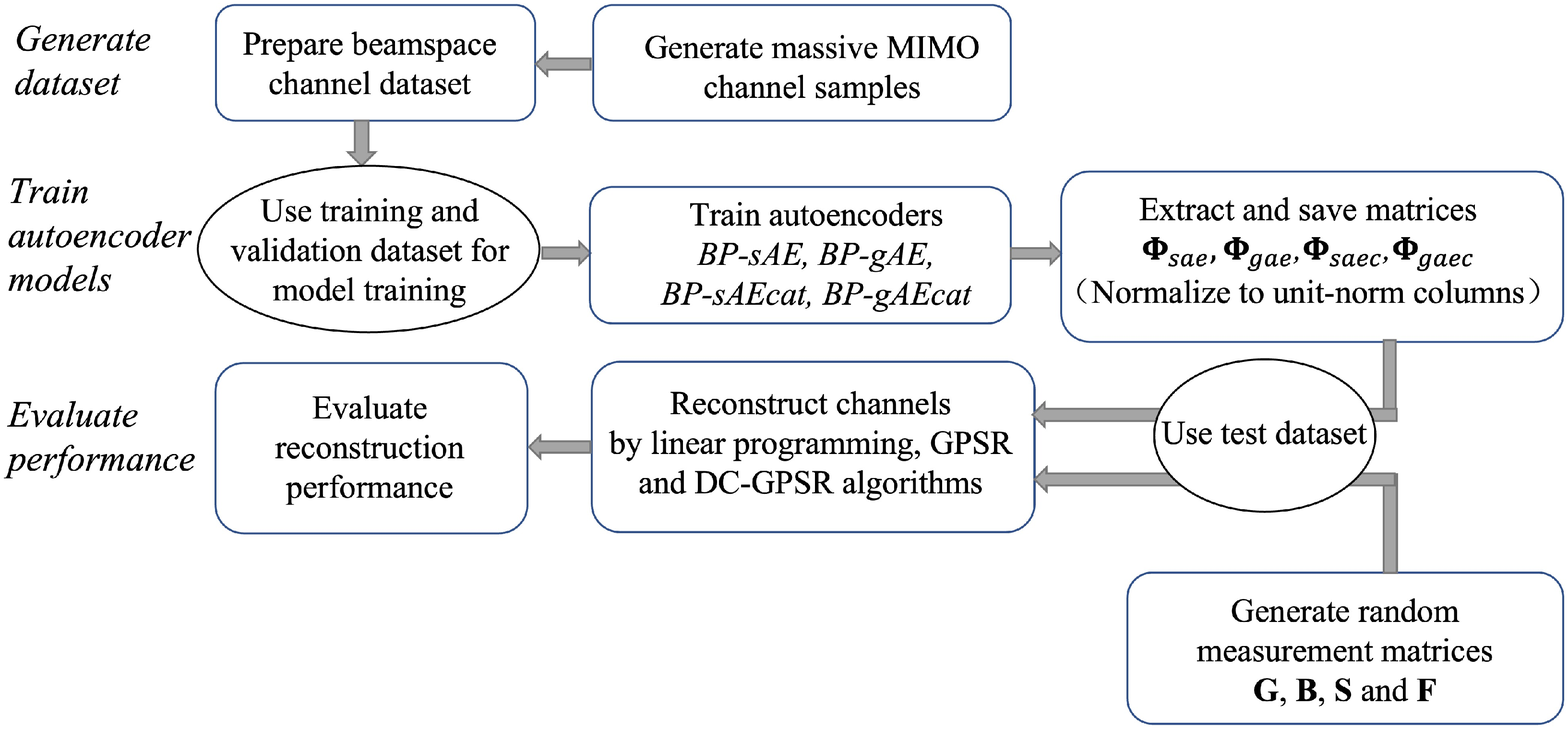}
	\caption{Diagram of the performed experiment}
	\label{exp_diag}
	\end{figure*}

	The performed experiment \footnote{\bb{Our experiment implementation is open to public at \url{https://github.com/pengxiawu/DeepBP-AE}.}} contains three phases including dataset generation, model training and performance evaluation, as shown in the \mbox{Fig. \ref{exp_diag}} diagram.
	In the phase of dataset generation, we randomly generate $50,000$ spatial-domain channel realizations according to the channel model in \eqref{spatial channel vector}.
	We consider a massive MIMO system having $256$ antennas at the BS and a single antenna at each of the UEs. 
	We generate spatial channels for each user and obtain a dataset containing spatial-domain channel vectors.
	For each channel, three scattering clusters were selected, including one line-of-sight path and two non-line-of-sight paths.
	The Ricean $K$-factor of the scattering clusters is set to be $13.2$ dB \cite{Zhu2017Auxiliary}.
	The AoDs of scattering clusters follow a uniform distribution between $[-\pi/2, \pi/2]$, and the complex path gains follow zero-mean Gaussian distribution.
	The spatial-domain channel vectors are transformed into \mbox{$S$-sparse} beamspace channel vectors with the sparsity level $S=16$ since it was observed that the $16$ largest-magnitude channel components are sufficient to occupy most of the channel energy.
	We stack the real part and imaginary part of each beamspace channel vector column-wise, and obtain a dataset $\{ \mathbf X\}$ consisting of $100,000$ \mbox{real-form} sparse vectors, which are $\{\mathbf X \}=\{ \left[\mathbf H_1, \mathbf H_{2},...\right] \} = \{ \left[\Re(\mathbf h_{b,1}), \Im(\mathbf h_{b,1}), \Re(\mathbf h_{b,2}), \Im(\mathbf h_{b,2}), ... \right]\}$.
	We normalize the channel vectors such that the performance is independent of the path loss.
	For simplicity, we will use the column vector $\mathbf x \in \mathbb{R}^{256}$ to represent an arbitrary sample in the dataset $\{\mathbf X\}$. 
	We split the samples of the dataset into training, validation and testing datasets by the ratio of $0.96/0.02/0.02$.
	The training and validation datasets are used to train the autoencoder models, and the testing dataset will be used for performance evaluations.
	The decoder depth is set to be $15$ layers, i.e., $L=15$. 
	The measurement matrix is initialized with a truncated normal distribution that has a standard deviation of \mbox{$\sigma = 1/\sqrt{256}$}. 
	\bb{The parameter setup in the model training phase is summarized in Table \ref{training}.}
	The training processes are repeated over different compressed dimensions between $24 \le M \le 72$.
	 After training concludes, we extract the weights as the data-driven measurement matrices $\mathbf \Phi_{sae}$, $\mathbf \Phi_{gae}$, $\mathbf \Phi_{saec}$ and $\mathbf \Phi_{gaec}$ from the trained autoencoder models $BP\textit{-}sAE$, $BP\textit{-}gAE$, $BP\textit{-}sAEcat$ and $BP\textit{-}gAEcat$, respectively.
	 
	
	\begin{table*}[t!]
  \begin{center}
  \captionsetup{justification=centering}
    \caption{Training Parameters}
    \label{training}
    \begin{tabular}{|c|c|c|c|c|c|} 
    \hline
    Training optimizer &  Step size $\alpha$ initialization & Learning rate  & Batch size & Maximum epochs & Early stopping patience\\
    \hline
      SGD \tablefootnote{\bb{Other advanced training algorithms, such as the Adam, can potentially improve the convergence and reduce the training time.}} & $1.0$ & $0.001$ & $128$ & $1,000$ epochs & $25$ epochs \\
    \hline
    \end{tabular}
  \end{center}
\end{table*}

	In the performance evaluation phase, we evaluate the reconstruction performance by applying the learned data-driven measurement matrices to classical sparse reconstruction algorithms, and the results will be presented in the following sections.
	The trained autoencoders can be used to perform reconstructions directly and therefore are used to perform sparse channel reconstructions over the testing dataset. 
	The testing normalized mean squared error \mbox{(NMSE) \footnote{The NMSE is defined as $\mathbb{E}\left[\norm{\mathbf{x}-\hat{\mathbf x}}_2 ^2 / \norm{\mathbf{x}}_2^2\right]$, where $\mathbf{x}$ is a sample from the testing dataset.}} of the proposed autoencoder models for different compressed dimensions $M$ are shown in Table \ref{test_error_ae}.
	We can observe that the test errors decrease when the compressed dimension $M$ increases. The $BP\textit{-}gAE$ has lower test errors among the four types of autoencoders.
	This result confirms that the proposed residual learning module for $BP\textit{-}gAE$, i.e., the term $\mathbf \Phi ^{T} \mathbf \Phi \mathbf x^{(t-1)}$ in \eqref{l1gae_decoder}, can improve data fitting compared with the $BP\textit{-}sAE$ model. 
	The $BP\textit{-}gAE$ has a test error below $0.01$ for $M \ge 64$; therefore, it is valid to perform sparse reconstructions
using the trained autoencoder $BP\textit{-}gAE$ when the measurements are sufficient.
	However, it should be noted that the $BP\textit{-}AE$ models are designed to acquire data-driven measurement matrices, instead of to be used for sparse reconstructions. 
	Therefore,  instead of using random matrices, we propose to apply the learned measurement matrices to the classical sparse reconstruction algorithms such that improved reconstructions can be achieved.
	The performance of the corresponding reconstructions are displayed and discussed in the following sections.
	

\begin{table*}[t!]
  \begin{center}
  \captionsetup{justification=centering}
    \caption{Reconstruction NMSE Over Testing Dataset of the Proposed Autoencoder Models}
    \label{test_error_ae}
    \begin{tabular}{c|c|c|c|c} 
    \hline
    \diagbox{$M$}{Test NMSE}{Models} & $BP\textit{-}sAE$ &  $BP\textit{-}gAE$ & $BP\textit{-}sAEcat$  & $BP\textit{-}gAEcat$ \\
      \hline
      24 & 0.201 & 0.147 & 0.047 & 0.047  \\
       \hline
      32 & 0.143 & 0.064 & 0.037 & 0.039 \\
        \hline
      40 & 0.104 & 0.030 & 0.038 & 0.034 \\
        \hline
      48 & 0.084 & 0.020 & 0.027 & 0.031 \\
        \hline
      56 & 0.057 & 0.012 & 0.027 & 0.024 \\
        \hline
      64 & 0.057 & \circled{0.009} & 0.024 & 0.022 \\
        \hline
      72 & 0.041 & \circled{0.008} & 0.022 & 0.022 \\
        \hline
    \end{tabular}
  \end{center}
\end{table*}

\subsection{Reconstructions by Linear Programming Recovery}
	In this section, we will evaluate reconstruction performance by linear programming recovery using various measurement matrices over the testing dataset.
	We denote $\mathbf \Phi_{sae} \in \mathbb{R}^{M\times 256}$, $\mathbf \Phi_{gae} \in \mathbb{R}^{M\times 256}$, $\mathbf \Phi_{saec} \in \mathbb{R}^{M\times 512}$ and $\mathbf \Phi_{gaec} \in \mathbb{R}^{M\times 512}$ as the learned data-driven matrices by $BP\textit{-}sAE$, $BP\textit{-}gAE$, $BP\textit{-}sAEcat$ and \mbox{$BP\textit{-}gAEcat$} respectively.	
	We choose four types of random matrices for comparisons including the random Gaussian matrix, the random Bernoulli matrix \footnote{For a random Bernoulli matrix, entries are $-1$ or $1$ with equal probability.}, the partial Fourier matrix and the random selection matrix \footnote{For a random selection matrix, entries are $0$ or $1$ with equal probability.}. 
	For the learned matrices $\mathbf \Phi_{sae}$, $\mathbf \Phi_{gae}$ and the random Gaussian matrix $\mathbf G \in \mathbb{R}^{M\times 256}$, the random Bernoulli matrix $\mathbf B \in \mathbb{R}^{M\times 256}$, the random selection matrix $\mathbf S \in \mathbb{R}^{M\times 256}$ and the partial Fourier matrix $\mathbf F \in \mathbb{R}^{M\times 256}$, we use the linear programming method to solve the sparse reconstruction problem in \eqref{bp}.
	For the learned matrices $\mathbf \Phi_{saec}$, $\mathbf \Phi_{gaec}$ and the random Gaussian matrix $\mathbf G_c \in \mathbb{R}^{M\times 512}$, random Bernoulli matrix $\mathbf B_c \in \mathbb{R}^{M\times 512}$, random selection matrix $\mathbf S_c \in \mathbb{R}^{M\times 512}$ and the partial Fourier matrix $\mathbf F_c \in \mathbb{R}^{M\times 512}$, we use linear programming with an auxiliary nonnegativity constraint to solve the sparse reconstruction problem in \eqref{norm1_minimization_cat}.

\begin{table*}[!t]
\centering
\captionsetup{justification=centering}
\caption {Reconstruction NMSE by Linear Programming Recovery with Different Compressed Dimensions $M$ for Various Measurement Matrices} 
\label{performance_nmse} 
\begin{tabular}{ l|c|c|c|c|c|c|c}
\hline
\diagbox[width=4.8cm]{Matrix type}{Reconstruction \\  NMSE}{$M$}  
& $M=24$  & $M=32$  & $M=40$  & $M=48$  & $M=56$ & $M=64$ & $M=72$  \\
\hline 
Learned matrix $\mathbf \Phi_{sae} \in \mathbb{R}^{M \times 256}$   
& $\mathbf{0.07}$ & $\mathbf{0.01}$ & $\mathbf{0.006}$ & $\mathbf{0.003}$ & $\mathbf{0.0006}$ & $\mathbf{0.0005}$ & $\mathbf{9 \times 10^{-5}}$  \\
\hline
Learned matrix $\mathbf \Phi_{gae}\in \mathbb{R}^{M \times 256}$  
& $\mathbf{0.04}$ & $\mathbf{0.01}$ & $\mathbf{0.003}$ & $\mathbf{0.001}$ & $\mathbf{0.0005}$ & $\mathbf{7 \times 10^{-6}}$ & $\mathbf{2 \times 10^{-6}}$  \\
\hline
Random Gaussian $\mathbf G \in \mathbb{R}^{M \times 256}$        
& $0.2$ & $0.1$ & $0.07$ & $0.04$ & $0.02$ & $0.006$ & $0.0006$   \\
\hline 
Random Bernoulli $\mathbf B\in \mathbb{R}^{M \times 256}$          
& $0.2$ & $0.1$ & $0.06$ & $0.04$ & $0.02$ & $0.005$ & $0.0005$    \\
\hline 
Partial Fourier $\mathbf F\in \mathbb{R}^{M \times 256}$           
& $0.4$ & $0.1$ & $0.06$ & $0.02$ & $0.07$ & $0.07$ & $0.02$   \\
\hline 
Random Selection $\mathbf S\in \mathbb{R}^{M \times 256}$        
& $0.2$ & $0.1$ & $0.06$ & $0.04$ & $0.02$ & $0.005$ & $0.0004$  \\
\hline 
\hline
Learned matrix $\mathbf \Phi_{saec} \in \mathbb{R}^{M \times 512}$  
& $\mathbf{0.1}$ & $\mathbf{0.06}$ & $\mathbf{0.03}$ & $\mathbf{0.004}$ & $\mathbf{0.001}$ & $\mathbf{3 \times 10^{-15}}$ & $\mathbf{3 \times 10^{-15}}$  \\
\hline
Learned matrix $\mathbf \Phi_{gaec} \in \mathbb{R}^{M \times 512}$  
& $\mathbf{0.01}$ & $\mathbf{0.004}$ & $\mathbf{0.0006}$ & $\mathbf{5\times 10^{-5}}$ & $\mathbf{5 \times 10^{-30}}$ & $\mathbf{5 \times 10^{-30}}$ & $\mathbf{6 \times 10^{-30}}$    \\ 
\hline
Random Gaussian $\mathbf G_c \in \mathbb{R}^{M \times 512} $        
& $0.2$ & $0.1$ & $0.07$ & $0.04$ & $0.02$ & $0.006$ & $0.0009$    \\
\hline 
Random Bernoulli $\mathbf B_c \in \mathbb{R}^{M \times 512}$          
& $0.2$ & $0.1$ & $0.07$ & $0.04$ & $0.02$ & $0.004$ & $0.0005$    \\
\hline 
Partial Fourier $\mathbf F_c \in \mathbb{R}^{M \times 512}$           
& $0.09$ & $0.1$ & $0.07$ & $0.05$ & $0.06$ & $1 \times 10^{-6}$ & $0.02$  \\
\hline 
Random Selection $\mathbf S_c \in \mathbb{R}^{M \times 512}$        
& $0.3$ & $0.2$ & $0.09$ & $0.05$ & $0.02$ & $0.006$ & $0.0007$  \\
\hline
\end{tabular}
\end{table*}

\begin{table*}[!t]
\centering
\captionsetup{justification=centering}
\caption {Accurate-Reconstruction Percentages by Linear Programming Recovery with Different Compressed Dimensions $M$ for Various Measurement Matrices} 
\label{performance_table2} 
\begin{tabular}{ l|c|c|c|c|c|c|c}
\hline
\diagbox[width=4.8cm]{Matrix type}{Accurate reconstruct \\ percentage ($\%$)}{$M$}  
& $M=24$  & $M=32$  & $M=40$  & $M=48$  & $M=56$ & $M=64$ & $M=72$    \\
\hline 
Learned matrix $\mathbf \Phi_{sae} \in \mathbb{R}^{M \times 256}$   
& $\mathbf{11.2}$ & $\mathbf{44.2}$ & $\mathbf{76.6}$ & $\mathbf{89}$ & $\mathbf{97.4}$ & $\mathbf{99.3}$ & $\mathbf{99.6}$  \\
\hline
Learned matrix $\mathbf \Phi_{gae} \in \mathbb{R}^{M \times 256}$  
& $\mathbf{11.9}$ & $\mathbf{53.8}$ & $\mathbf{78.5}$ & $\mathbf{92.5}$ & $\mathbf{98}$ & $\mathbf{99.8}$ & $\mathbf{99.9}$  \\
\hline
Random Gaussian $\mathbf G  \in \mathbb{R}^{M \times 256}$        
& $0$ & $0$ & $0.1$ & $1.6$ & $32.1$ & $61.9$ & $95.7$   \\
\hline 
Random Bernoulli $\mathbf B \in \mathbb{R}^{M \times 256}$          
& $0$ & $0$ & $0$ & $3.3$ & $32.8$ & $70$ & $95.2$    \\
\hline 
Partial Fourier $\mathbf F \in \mathbb{R}^{M \times 256}$           
& $0.1$ & $0.1$ & $0.1$ & $16.7$ & $0.3$ & $0.0$ & $8.5$  \\
\hline 
Random Selection $\mathbf S \in \mathbb{R}^{M \times 256}$        
& $0$ & $0$ & $0.1$ & $4.6$ & $24.4$ & $74.7$ & $98.1$  \\
\hline \hline
Learned matrix $\mathbf \Phi_{saec} \in \mathbb{R}^{M \times 512}$  
& $\mathbf{16.9}$ & $\mathbf{61.9}$ & $\mathbf{87.3}$ & $\mathbf{98.9}$ & $\mathbf{99.9}$ & $\mathbf{100}$ & $\mathbf{100}$  \\
\hline
Learned matrix $\mathbf \Phi_{gaec} \in \mathbb{R}^{M \times 512}$  
& $\mathbf{21.9}$ & $\mathbf{60.1}$ & $\mathbf{90.7}$ & $\mathbf{98.1}$ & $\mathbf{100}$ & $\mathbf{100}$ & $\mathbf{100}$    \\ 
\hline
Random Gaussian $\mathbf G_c \in \mathbb{R}^{M \times 512} $        
& $0$ & $0$ & $0$ & $1.8$ & $21.8$ & $68.4$ & $92.4$   \\
\hline 
Random Bernoulli $\mathbf B_c \in \mathbb{R}^{M \times 512}$          
& $0$ & $0$ & $0$ & $1.9$ & $32.5$ & $77.7$ & $96$   \\
\hline 
Partial Fourier $\mathbf F_c \in \mathbb{R}^{M \times 512}$           
& $0$ & $0$ & $1$ & $0.7$ & $0.1$ & $99.8$ & $31.8$  \\
\hline 
Random Selection $\mathbf S_c \in \mathbb{R}^{M \times 512}$        
& $0$ & $0$ & $0$ & $1.6$ & $22.5$ & $67.5$ & $95.7$ \\
\hline
\end{tabular}
\end{table*}
    
	Table \ref{performance_nmse} shows the reconstruction NMSE for different matrices with different compressed dimensions $M$.
	The NMSE generally decreases when the compressed dimension $M$ increases for all matrices. 
	The learned data-driven matrices $\mathbf \Phi_{sae}$, $\mathbf \Phi_{gae}$, $\mathbf \Phi_{saec}$ and $\mathbf \Phi_{gaec}$ significantly outperform the random matrices for all compressed dimensions of $M$ in terms of reconstruction NMSE.
	The learned matrix $\mathbf \Phi_{gae}$ has a lower reconstruction NMSE than the learned matrix $\mathbf \Phi_{sae}$; the learned matrices $\mathbf \Phi_{saec}$ and $\mathbf \Phi_{gaec}$ have lower reconstruction errors than the learned matrices $\mathbf \Phi_{sae}$ and $\mathbf \Phi_{gae}$.
	The learned matrix $\mathbf \Phi_{gaec}$ can reconstruct accurately with an NMSE on the order of $10^{-30}$ when the compressed dimensions are $M=48$, $M=56$ and $M=72$.
	Table \ref{performance_table2} shows the accurate-reconstruction percentages, which refers to the ratio of the number of accurate reconstructed samples over the total number of samples in the testing dataset.
	We consider that a sample is accurately reconstructed when the normalized squared $\ell_2$-error of its reconstruction is no more than $10^{-8}$, i.e., $\norm {{\mathbf x} - \hat{\mathbf x}}_2^2 \le 10^{-8}$. 
	In the small compressed dimension range, such as for $24 \le M \le 40$, the random matrices cannot reconstruct accurately and show almost no accurate reconstructions, while the data-driven measurement matrices have significantly higher accurate-reconstruction percentages. 
	For example, the learned matrices $\mathbf \Phi_{gae}$, $\mathbf \Phi_{saec}$ and $\mathbf \Phi_{gaec}$ can achieve over $50\%$ accurate recoveries for a compressed dimension of $M=32$, whereas the random matrices have percentages which are almost all zeros. 
	When $M=40$ the data-driven matrix $\mathbf \Phi_{gaec}$ achieves over $90\%$ accurate reconstructions, while the percentages for the random matrices are all below $1 \%$. 
	From Table \ref{performance_nmse} and Table \ref{performance_table2}, we can infer that the learned matrices can achieve more accurate reconstructions using fewer measurements.
	At the same level of reconstruction accuracy, the learned data-driven matrices can reduce the requiring measurements by approximately $2S$, where $S=16$ represents the sparsity level.
	For example, in Table \ref{performance_nmse}, random matrices can achieve an NMSE on the order of $10^{-3}$ when $M=64$, while the learned matrix $\mathbf \Phi_{gaec}$ can achieve the same accuracy at $M=32$. 
 	In Table \ref{performance_table2}, the learned matrix $\mathbf \Phi_{gaec}$ can achieve a reconstruction percentage over $90\%$ at the compressed dimension $M=40$, while the required least number of measurements is $M=64$ for the random matrix $\mathbf F_c$ and is $M=72$ for the other types of random matrices.
%
%
	
	\begin{figure*}[!tb]  
	\centering
	\includegraphics[width=3.3in]{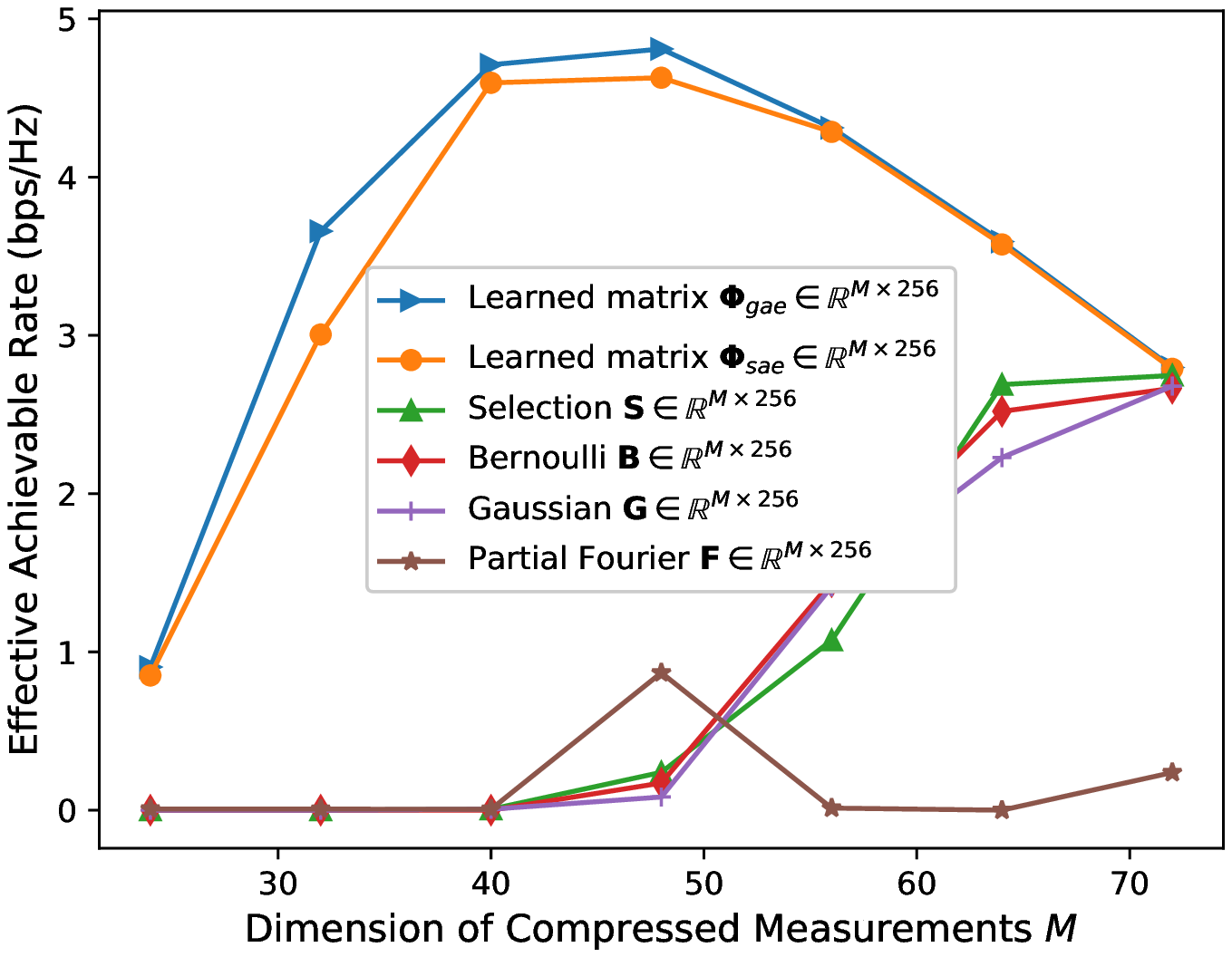}
	\includegraphics[width=3.3in]{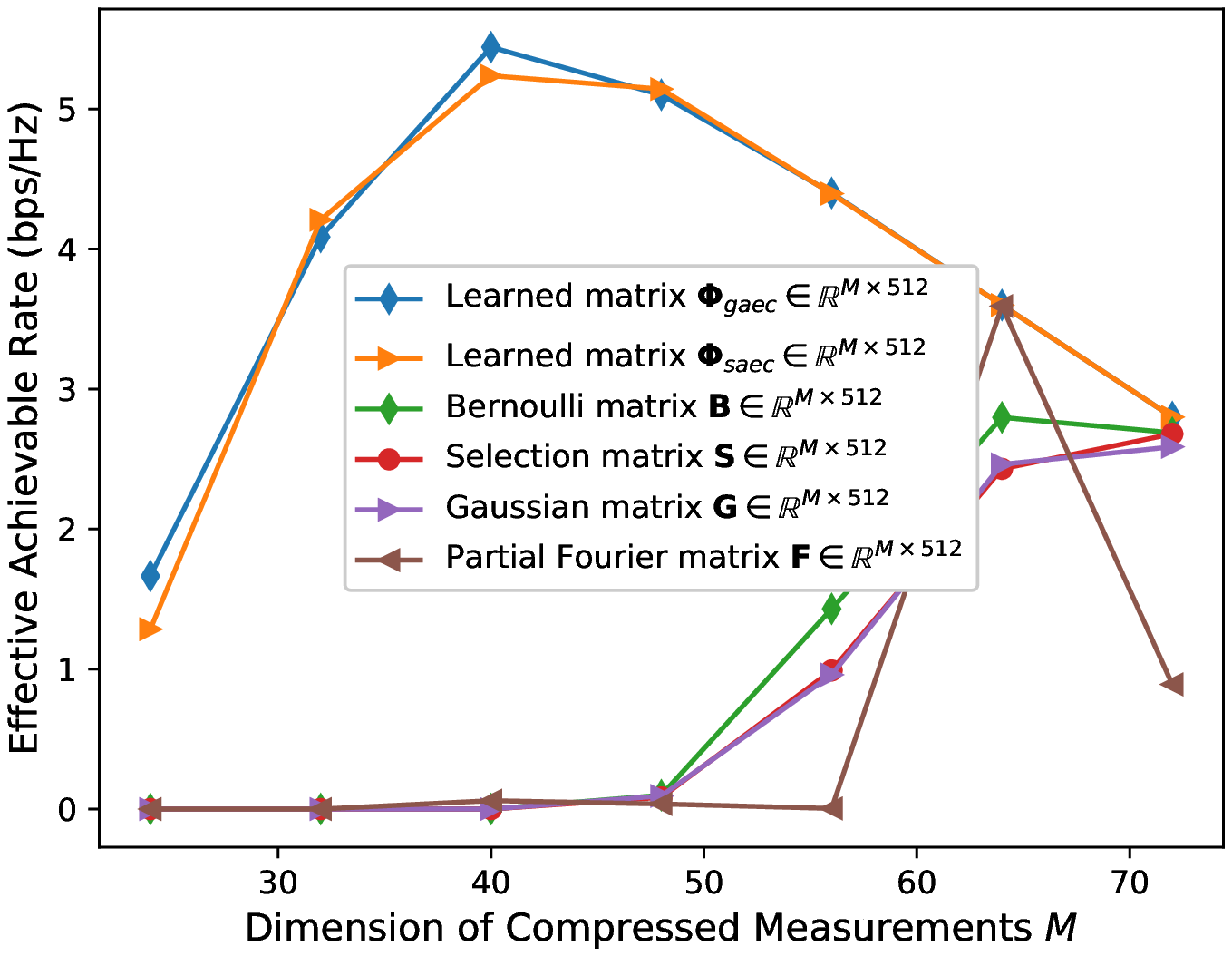}
	\caption{Effective achievable rates versus compressed dimension $M$}
	\label{sum_rate}
	\end{figure*}

    A larger compressed dimension $M$ will lead to a higher reconstruction accuracy, but a lower spectral efficiency.
    To analyze the tradeoff between compressed dimension $M$ and recovery accuracy, we define the effective achievable rate as $R_{e} = R_0 (1 - M / B) P_r$ \cite{alkhateeby2015compressed}, where $R_{0}$ is the average achievable rate for a user when perfect CSI is available and it is set as $R_{0}=10$, $M/B$ is the pilot occupation ratio in a transmission block, $B$ is the block length that is set as $100$ symbols and $P_r$ is the probability of successful recoveries.
    The effective achievable rates for various measurement matrices are shown in \mbox{Fig. \ref{sum_rate}}.
    The learned data-driven matrices significantly outperform random matrices in terms of effective achievable rates.
    In particular, the learned matrices $\mathbf \Phi_{sae}$ and $\mathbf \Phi_{gae}$ have higher effective achievable rates than random matrices $\mathbf G$, $\mathbf B$, $\mathbf F$ and $\mathbf S$, and the learned matrices $\mathbf \Phi_{saec}$ and $\mathbf \Phi_{gaec}$ have higher effective achievable rates than random matrices $\mathbf G_c$, $\mathbf B_c$, $\mathbf F_c$ and $\mathbf S_c$. However, it should be noted that the learned matrices $\mathbf \Phi_{saec}$ and $\mathbf \Phi_{gaec}$ have higher effective achievable rates than the learned matrices $\mathbf \Phi_{sae}$ and $\mathbf \Phi_{gae}$.
   	Moreover, the maximum effective achievable rates occur at smaller compressed dimensions $M$ for the learned matrices when compared with random matrices.

\subsection{Reconstructions Using the DC-GPSR Algorithm}

%

  	\begin{figure*}[!tb]  
	\centering
	\includegraphics[width=5.5in]{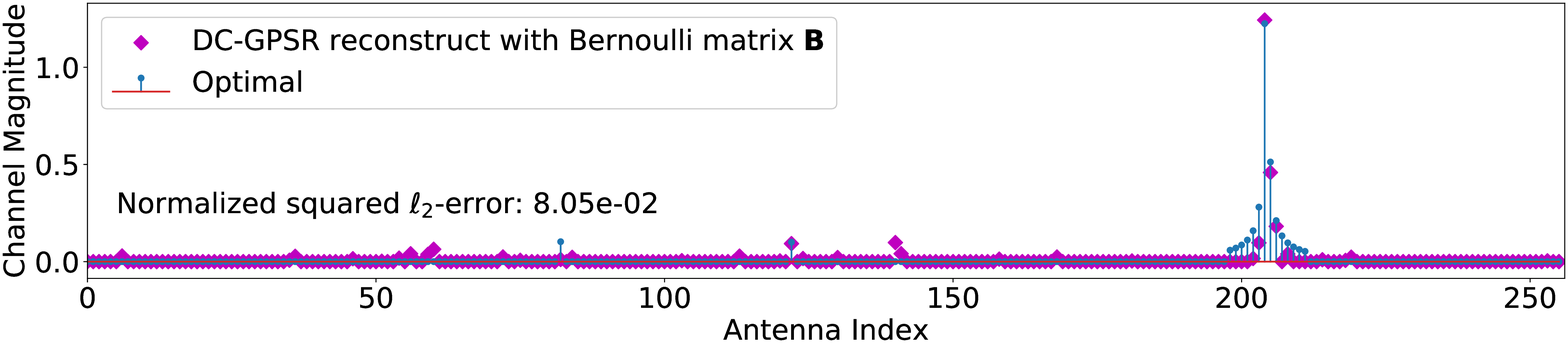}
	\includegraphics[width=5.5in]{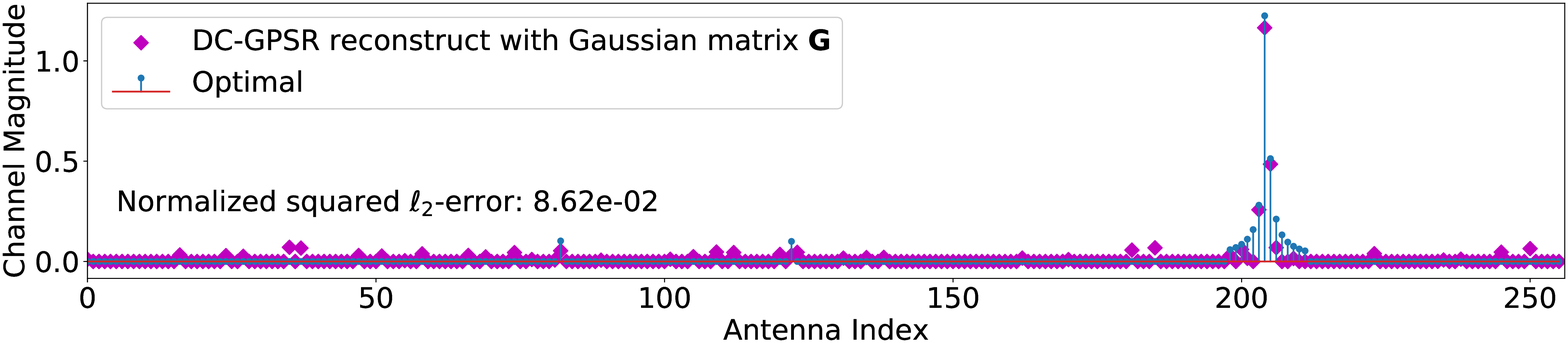}
	\includegraphics[width=5.5in]{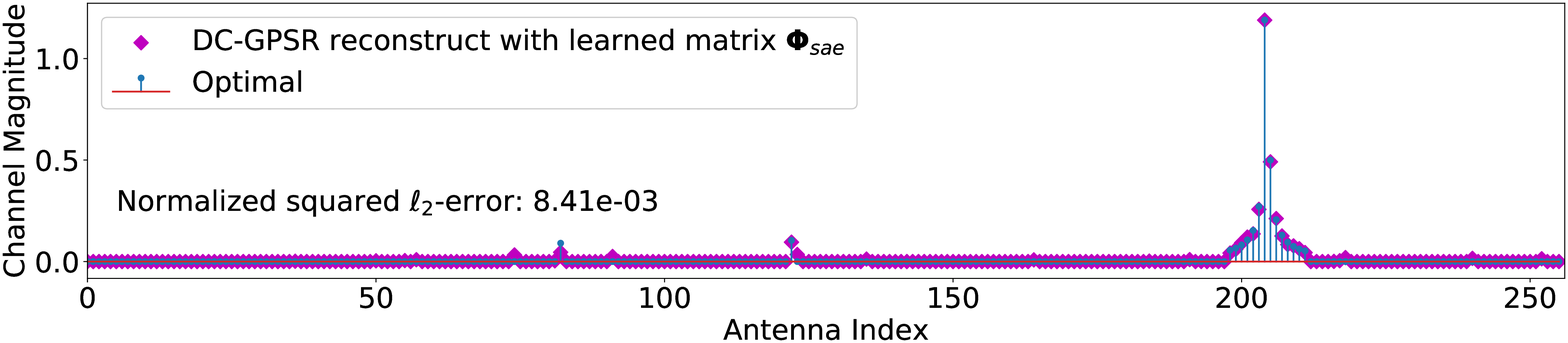}
	\includegraphics[width=5.5in]{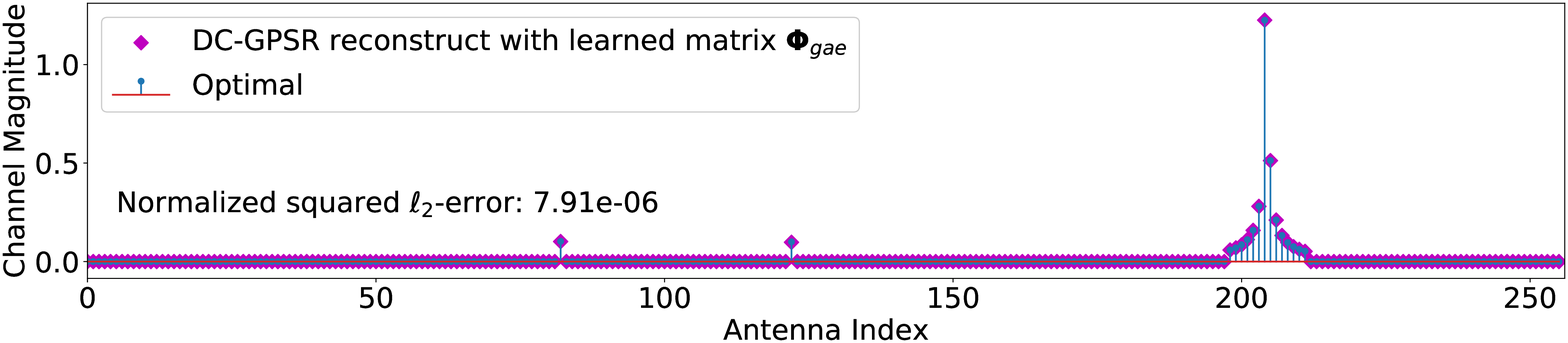}
	\captionsetup{justification=centering}
	\caption{Sparse channel reconstruction illustrations produced using the DC-GPSR algorithm with measurement matrices of the size $32\times 256$}
	\label{recon_dc_gpsr_emb32}
	\end{figure*}
	
	\begin{figure*}[!tb]  
	\centering
	\includegraphics[width=5.5in]{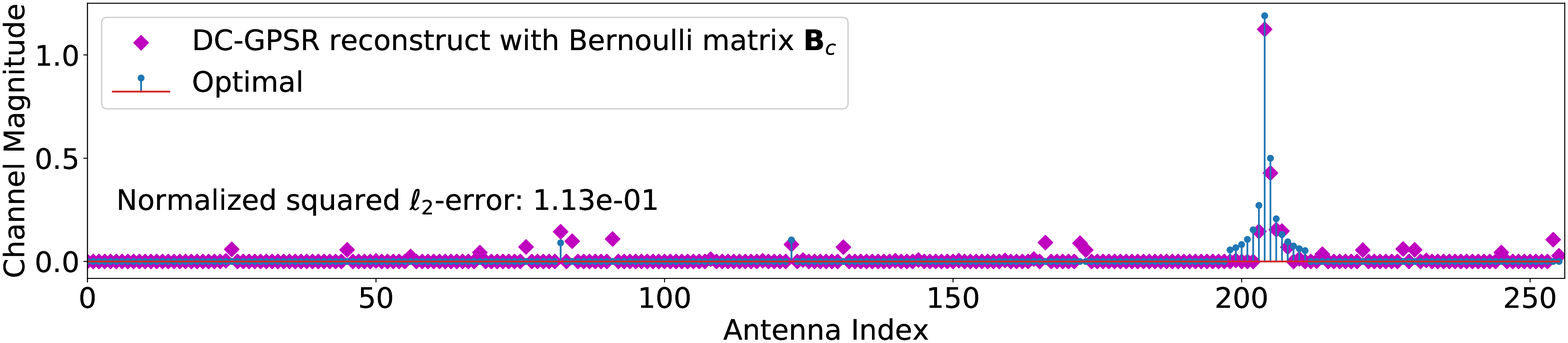}
	\includegraphics[width=5.5in]{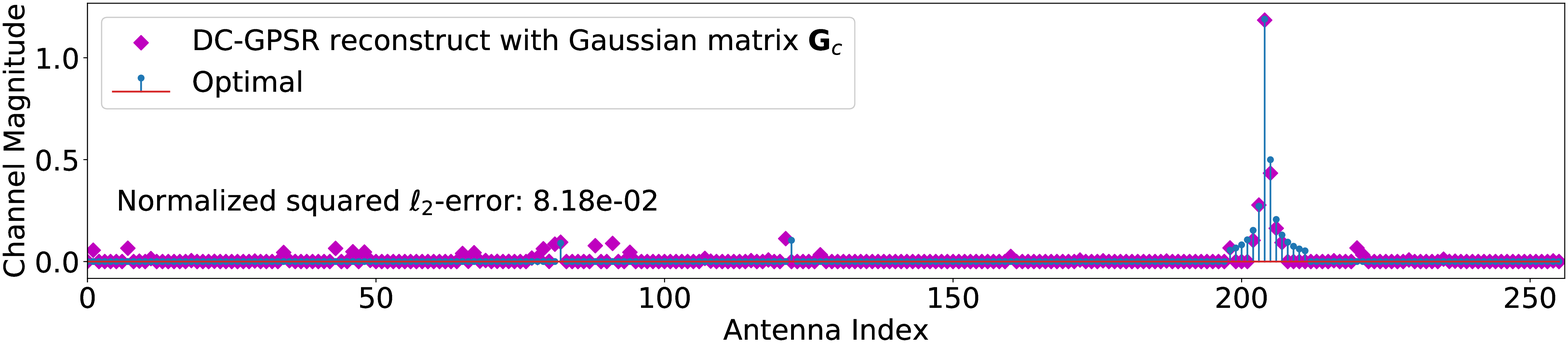}
	\includegraphics[width=5.5in]{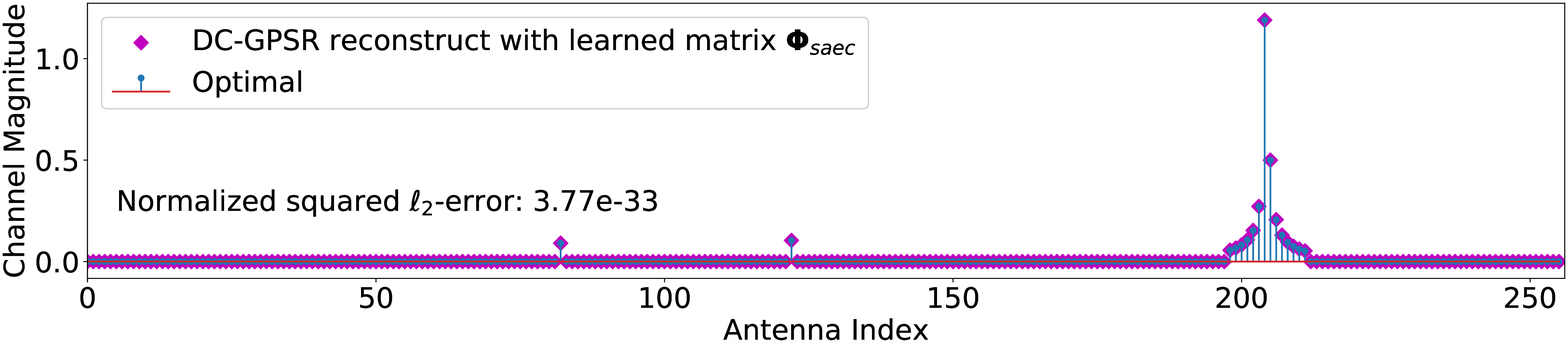}
	\includegraphics[width=5.5in]{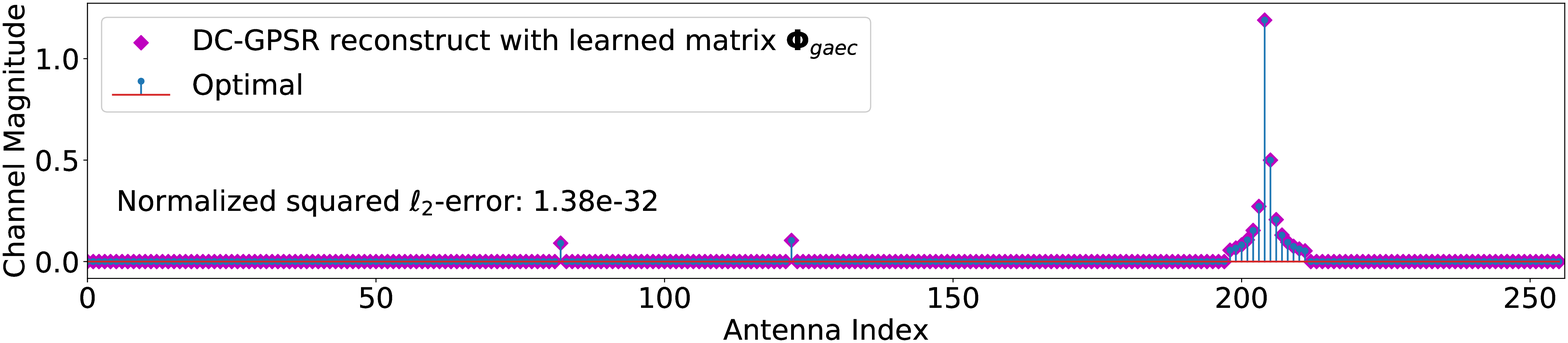}
	\captionsetup{justification=centering}
	\caption{Sparse channel reconstruction illustrations produced using the DC-GPSR algorithm with measurement matrices of the size $48 \times 512$}
	\label{recon_dc_gpsr_emb48}
	\end{figure*}
	
	The DC-GPSR algorithm was used to perform reconstructions with different measurement matrices after choosing a sample of a beamspace channel.
	In \mbox{Fig. \ref{recon_dc_gpsr_emb32}}, we plot the magnitudes of the original channel and reconstructed channels using the learned matrices $\mathbf \Phi_{sae} \in \mathbb{R}^{32 \times 256}$ and $\mathbf \Phi_{gae} \in \mathbb{R}^{32 \times 256}$, the random Gaussian matrix $\mathbf G \in \mathbb{R}^{32 \times 256}$ and random Bernoulli matrix $\mathbf B \in \mathbb{R}^{32 \times 256}$.
	The normalized squared $\ell_2$-error is on the order of $10^{-2}$ for the random Bernoulli and Gaussian matrices, while the errors are on the order of $10^{-3}$ for the learned matrix $\mathbf \Phi_{sae}$ and are on the order of $10^{-6}$ for the learned matrix $\mathbf \Phi_{gae}$.
	In Fig. \ref{recon_dc_gpsr_emb48}, we show the magnitudes of the original channel and reconstructed channels using the learned matrices $\mathbf \Phi_{saec} \in \mathbb{R}^{48 \times 512}$, $\mathbf \Phi_{gaec} \in \mathbb{R}^{48\times 512}$, the random Gaussian matrix $\mathbf G \in \mathbb{R}^{48 \times 512}$ and random Bernoulli matrix $\mathbf B \in \mathbb{R}^{48 \times 512}$.
	 The normalized squared \mbox{$\ell_2$-errors} are on the orders of $10^{-1}$ and $10^{-2}$ respectively for the random Bernoulli and Gaussian matrices, whereas the reconstructions are much more accurate using the learned matrices $\mathbf \Phi_{gaec}$ and $\mathbf \Phi_{saec}$, which achieve normalized squared $\ell_2$-errors on the order of $10^{-32}$ and $10^{-33}$.

\subsection{Reconstructions Using the GPSR Algorithm}

	\begin{figure*}[!tb]  
	\centering
	\subfloat[Compressed dimension $M=24$]
		{\includegraphics[width=3.3in]{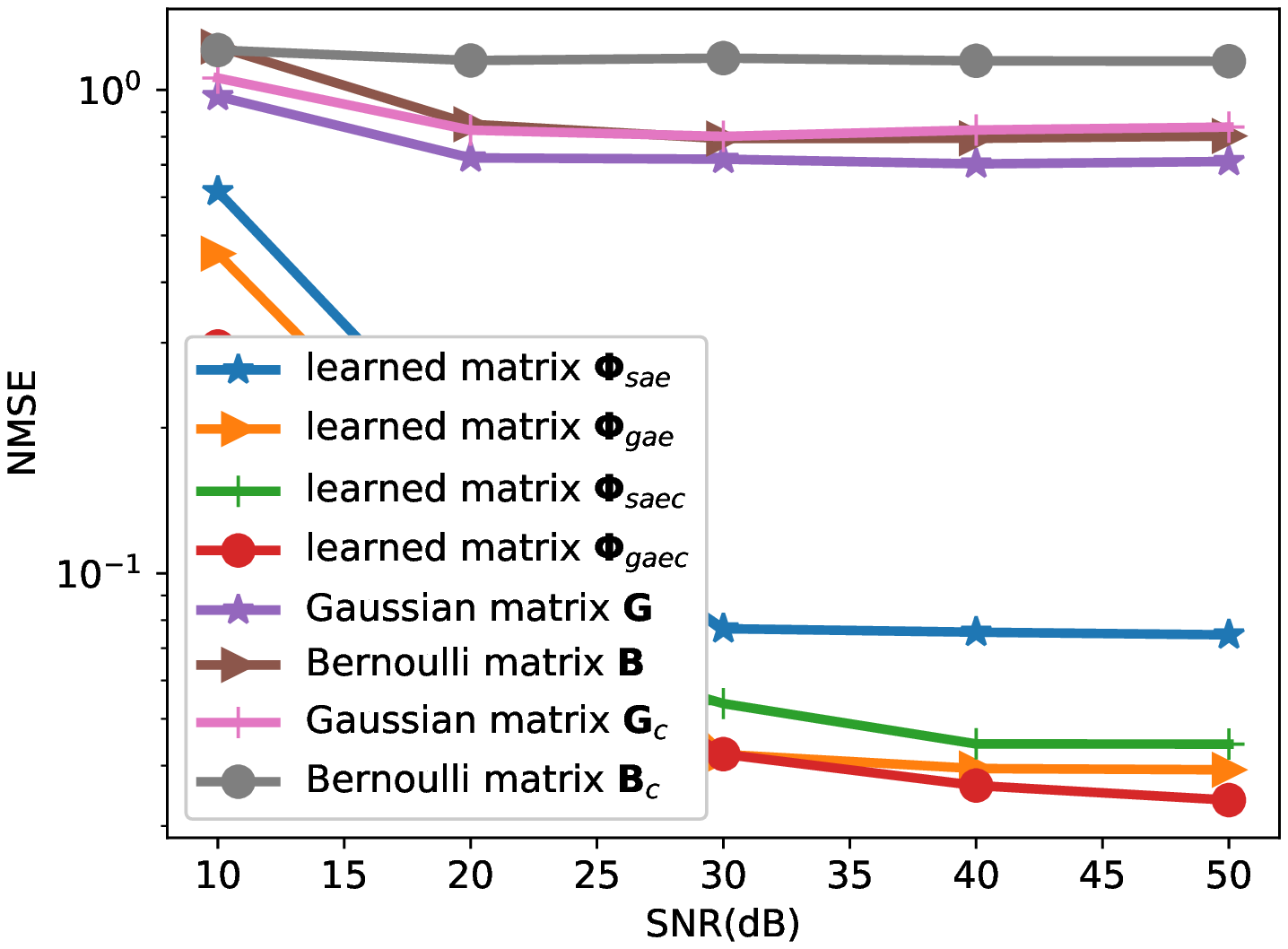}}
	\subfloat[Compressed dimension $M=40$]
		{\includegraphics[width=3.3in]{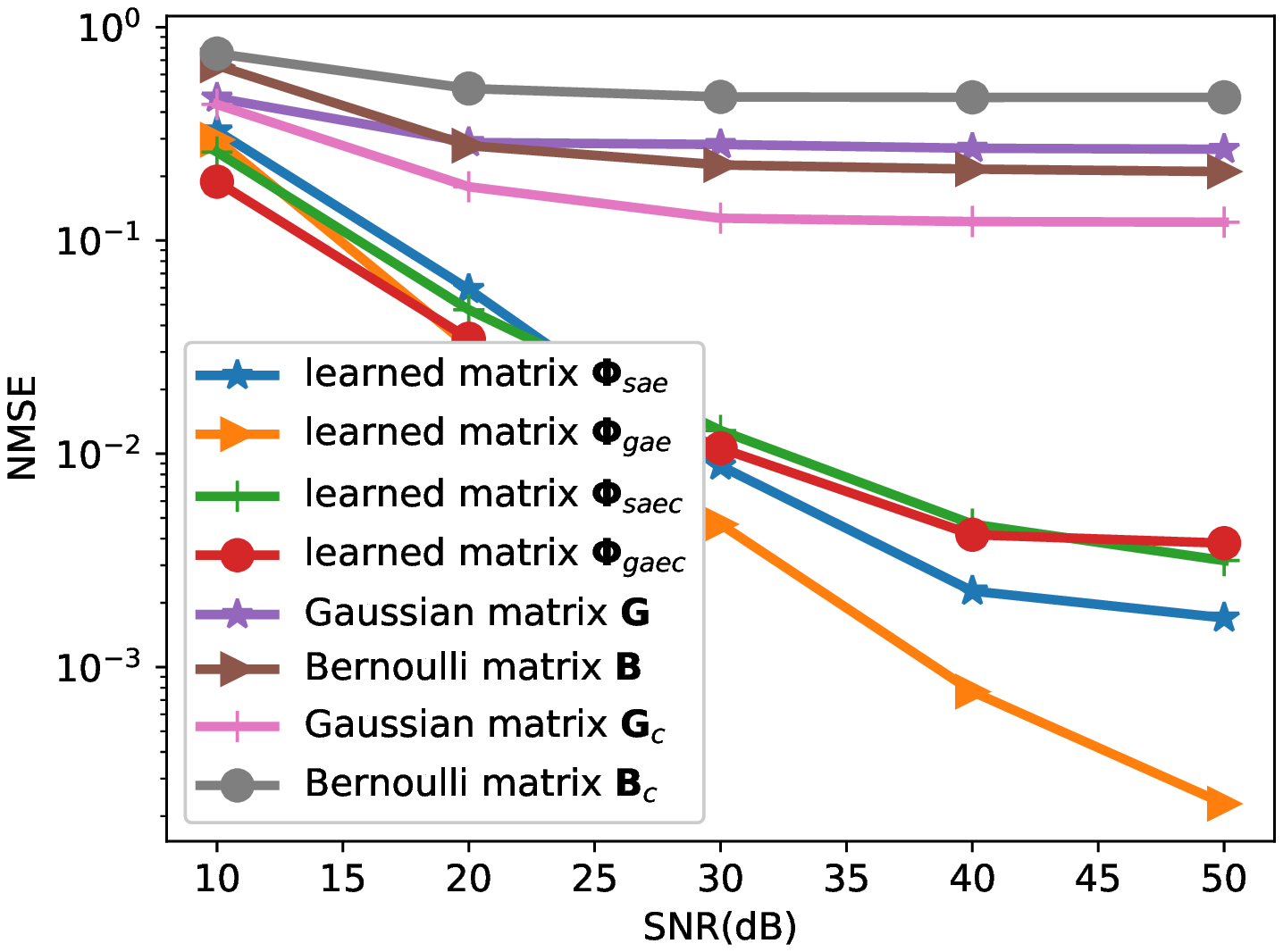}}\\
	\subfloat[Compressed dimension $M=56$]
		{\includegraphics[width=3.3in]{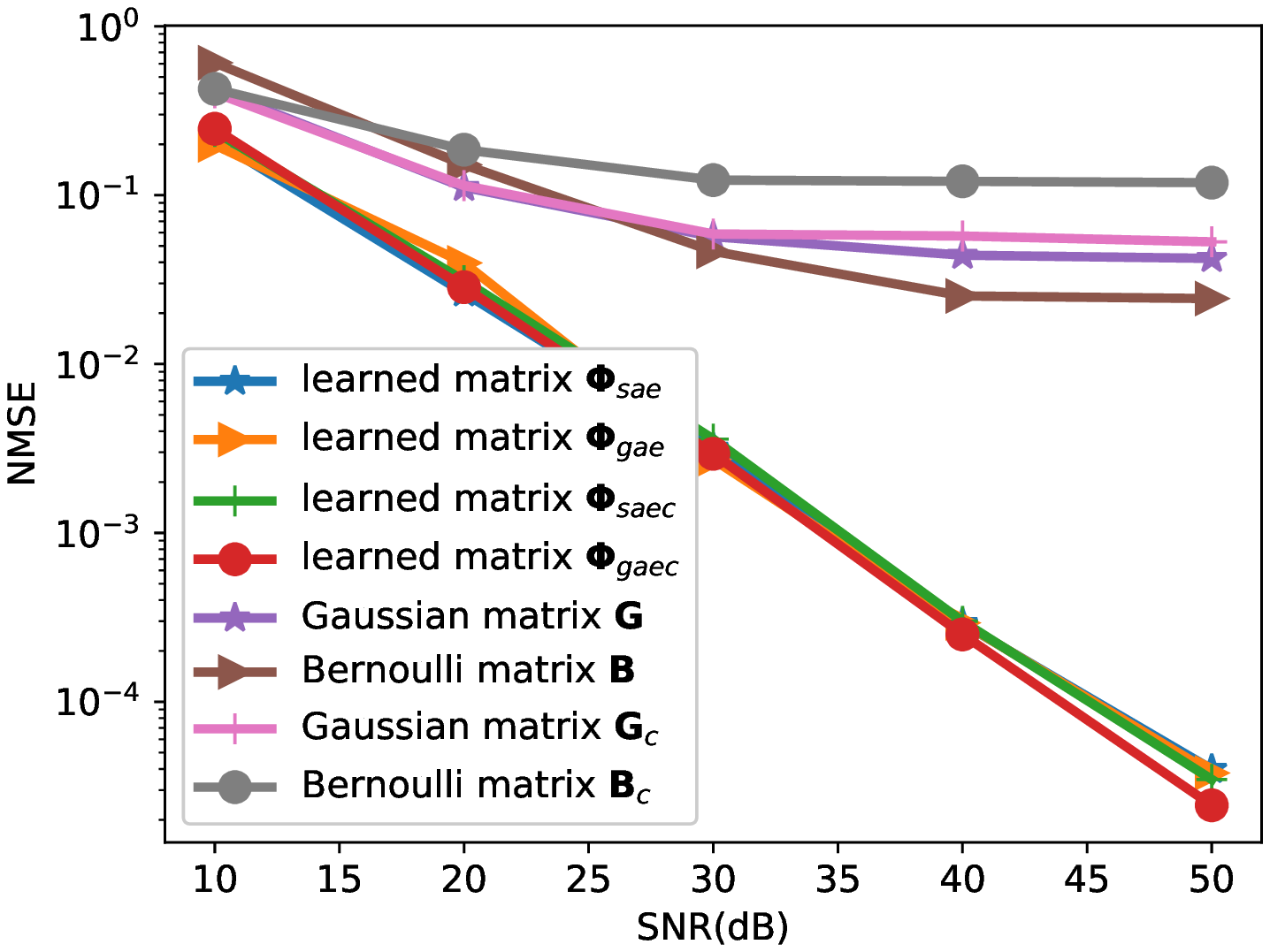}}
	\subfloat[Compressed dimension $M=72$]
		{\includegraphics[width=3.3in]{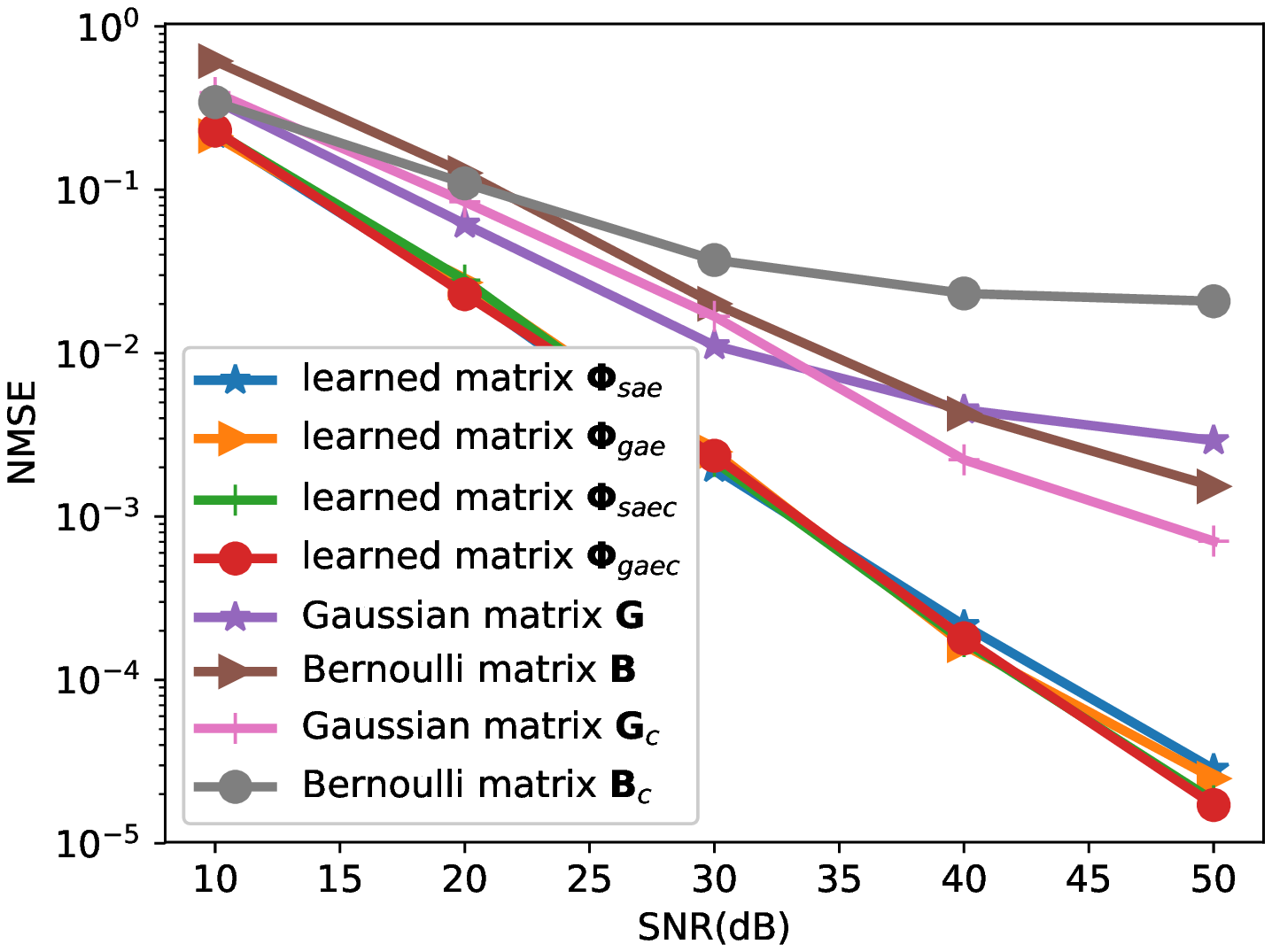}}
	\captionsetup{justification=centering}
	\caption{Reconstruction NMSE versus SNR by GPSR algorithm at compressed dimensions $M \in \{24, 40, 56, 72\}$ for the learned matrices $\mathbf \Phi_{sae} \in \mathbb{R}^{M \times 256}$, $\mathbf \Phi_{gae} \in \mathbb{R}^{M \times 256}$, $\mathbf \Phi_{saec} \in \mathbb{R}^{M \times 512}$, $\mathbf \Phi_{gaec} \in \mathbb{R}^{M \times 512}$ and random matrices $\mathbf G \in \mathbb{R}^{M \times 256}$, $\mathbf B \in \mathbb{R}^{M \times 256}$, $\mathbf G_{c} \in \mathbb{R}^{M \times 512}$, $\mathbf B_{c} \in \mathbb{R}^{M \times 512}$}
	\label{recon_gpsr_snr}
	\end{figure*}
 
 	The GPSR algorithm was adopted to perform reconstructions in noisy scenarios using $1,000$ channel samples. The results of NMSE versus signal-to-noise ratio (SNR) are plotted in \mbox{Fig. \ref{recon_gpsr_snr}} for the compressed dimensions $M \in {32, 40, 56, 72}$. The GPSR algorithm was proposed to solve the $\ell_1$-norm constrained LS optimization for sparse reconstructions \cite{figueiredo2007gradient}.
	From the four subplots in \mbox{Fig. \ref{recon_gpsr_snr}}, it is evident that the NMSE is generally more accurate at a larger compressed dimension for both the learned matrices and the random matrices.
	For all of the compressed dimensions, the learned matrices show lower reconstruction errors than the random matrices.
	As shown in \mbox{Fig. \ref{recon_gpsr_snr}} (a) and Fig. \ref{recon_gpsr_snr} (b), at small compressed dimensions such as $M=32$ and $M=40$, the learned matrices show distinct reconstructions errors, and the learned matrix $\mathbf \Phi_{gae}$ shows the best reconstruction accuracy.
	As shown in \mbox{Fig. \ref{recon_gpsr_snr}} (c) and Fig. \ref{recon_gpsr_snr} (d), at large compressed dimensions, such as $M=56$ and $M=72$, the learned matrices show similar reconstruction accuracy, and their NMSE plots coincide together.
	The random matrices can achieve lower reconstruction errors at larger compressed dimensions; but even at $M=72$, the random matrices have less accurate reconstructions than the learned \mbox{data-driven} matrices.

\subsection{Complexity-Accuracy Tradeoffs and Implementations of the Learned Matrices}
	We have proposed four autoencoder models to acquire different data-driven matrices.	
	\bb{It is worth mentioning several technical insights between the four autoencoder models.
	By comparing the forward-pass computations of the $BP$-$sAE$ and $BP$-$gAE$ in \eqref{l1sae_decoder} and \eqref{l1gae_decoder}, we conclude that the residual learning unit $\mathbf \Phi ^{T} \mathbf \Phi \mathbf x^{(t-1)} - \mathbf \Phi^{T} \mathbf \Phi \mathbf x^{(t)}$ can improve the measurement matrix optimization, since the reconstruction of the matrix $\mathbf \Phi_{gae}$ shows more noticeable improvements than the matrix $\mathbf \Phi_{sae}$.
	Furthermore, we observe the $\mathbf \Phi_{saec}$ and $\mathbf \Phi_{gaec}$ matrices acquired by the extension models $BP$-$sAEcat$ and $BP$-$gAEcat$ achieve more accurate reconstructions compared to the $\mathbf \Phi_{sae}$ and $\mathbf \Phi_{gae}$ matrices.
	Considering the matrix sizes, i.e., the $\mathbf \Phi_{saec}$ and $\mathbf \Phi_{gaec}$ matrices have the size of $M \times 2N$ while the $\mathbf \Phi_{sae}$ and $\mathbf \Phi_{gae}$ matrices have the size of $M \times N$, we notice it is an interesting observation since intuitively one tends to believe that wider matrices having more columns can degrade the reconstruction performance due to a larger compression ratio.
	For example, the random matrices in the size of $M \times 2N$ show degraded performance compared to the corresponding random matrices in the size of $M \times N$, as shown in Table II and Table III.
	On the contrary, the $\mathbf \Phi_{saec}$ and $\mathbf \Phi_{gaec}$ matrices show improved performance. 
	One plausible reason is that the larger-size matrices $\mathbf \Phi_{saec}$ and $\mathbf \Phi_{gaec}$ have more variables to be optimized in the neural network, which means more degrees of freedom.
	Another insight is that the introduced nonnegativity constraint $\tilde{\mathbf x} \ge \mathbf{0}$ in \eqref{norm1_minimization_cat} for the extension models $BP$-$sAEcat$ and $BP$-$gAEcat$ provides an auxiliary feature that can be easily exploited by the neural network, leading to helpful adaptation to this auxiliary feature for the optimizing measurement matrices.}
	
	In practical applications, we need to consider the tradeoffs between the computational complexity and the reconstruction performance to choose an appropriate data-driven measurement matrix.
	The training complexities for acquiring these four types of data-driven matrices are $\mathbf \Phi_{sae} < \mathbf \Phi_{gae} < \mathbf \Phi_{saec} < \mathbf \Phi_{gaec}$.
	Apart from training complexity, another issue to be considered is the computational complexity in compressive sensing reconstruction algorithms. 
	For linear programming recovery, the matrices $\mathbf \Phi_{saec}$ and $\mathbf \Phi_{gaec}$ are twice as difficult to compute than the matrices $\mathbf \Phi_{sae}$ and $\mathbf \Phi_{gae}$ since the optimization problem requires solving dimensions that are twice as big.
	However, when we use the GPSR and DC-GPSR algorithms for reconstructions, the computational complexities for the four types of learned matrices are the same.
	We also should consider the dimension limitations when using the matrices in specific applications.
	For example, the downlink pilot matrix design requires the matrix size to be $M \times N$ to match the unknown channel vectors of the dimension $N$, whereas the CSI feedback compression can use matrices having sizes of both $M \times N$ and $M \times 2N$, since the UE can transform the known beamspace channel vectors into \mbox{$\tilde{\mathbf h}_b= [(\mathbf h)_{b,+}^T, (-\mathbf h)_{b,+}^T]^T$} of the dimension $2N$; then we can use a measurement matrix of the size $M \times 2N$ to compress the channel vectors.
	Essentially, the matrices $\mathbf \Phi_{saec}$ and $\mathbf \Phi_{gaec}$ are suggested to be used in CSI feedback and the matrices $\mathbf \Phi_{sae}$ and $\mathbf \Phi_{gae}$ are suitable to be used for downlink pilot designs.


\vspace{-4mm}
\subsection{Comparisons With Other Deep Learning Methods}
	\bb{The proposed measurement matrix design leads to a hybrid data-driven approach for sparse channel reconstructions that applies data-driven measurement matrices in traditional sparse reconstruction algorithms. It is interesting to compare the performance of the hybrid data-driven sparse reconstructions with the pure deep learning sparse reconstruction methods. 
	We choose a model-based autoencoder with the LISTA decoder \cite{Gregor2010Learning, Monga2021algorithm}, and a model-free autoencoder with a CNN decoder as the benchmarks.
	The LISTA based autoencoder is implemented by changing the decoder in our proposed autoencoder framework to the LISTA deep network. 
	The CNN method is set up according to the proposed implementation \cite{Ma2020data}, which is a DNN consisting of a linear encoder and stacked CNN decoder. 
	We use the same dataset to train the LISTA and CNN-based autoencoders and then use the trained network to perform sparse reconstructions.
	To implement the proposed hybrid data-driven method, we adopt the DC-GPSR algorithm to perform sparse reconstructions over the testing dataset with the proposed data-driven measurement matrices.
	The NMSE results versus different noise levels for compressed dimensions $M$ = 48 and $M$ = 64 are shown in Fig. \ref{exp_diag}, where $\sigma$ is the standard deviation of the Gaussian noise added on the measurements.
	We can see that the proposed data-driven measurement matrices with DC-GPSR reconstruction algorithm can achieve lower NMSE than the LISTA and CNN-based autoencoders. When the noise level is low, the proposed schemes has noticeably higher accuracy compared with the LISTA and CNN-based autoencoders.}  
	
	\begin{figure*}[!tb]
	\centering
		\captionsetup{justification=centering}
	\subfloat[Compressed dimension $M=48$]
	{\includegraphics[scale=0.50]{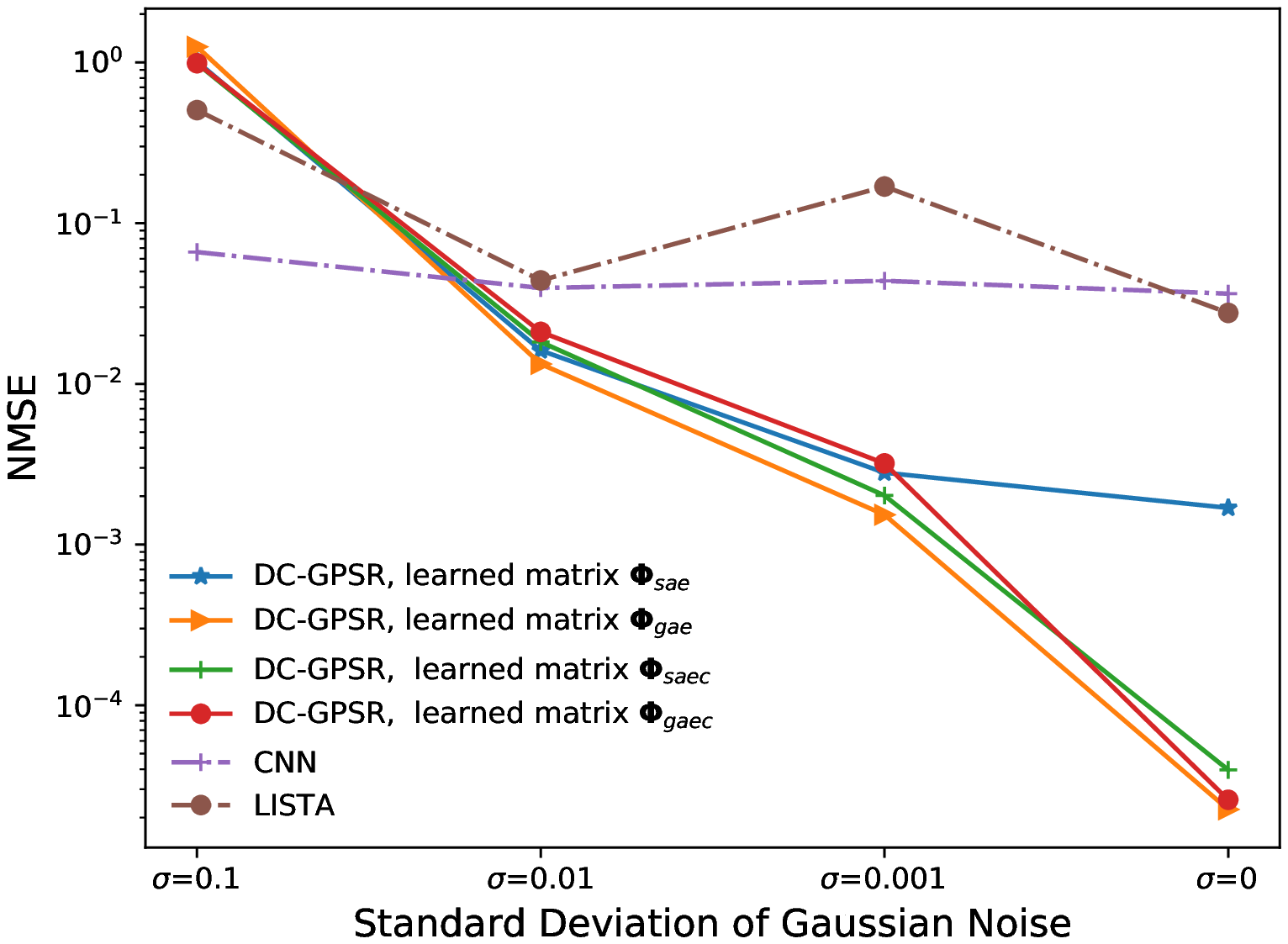}}
	\subfloat[Compressed dimension $M=64$]
	{\includegraphics[scale=0.50]{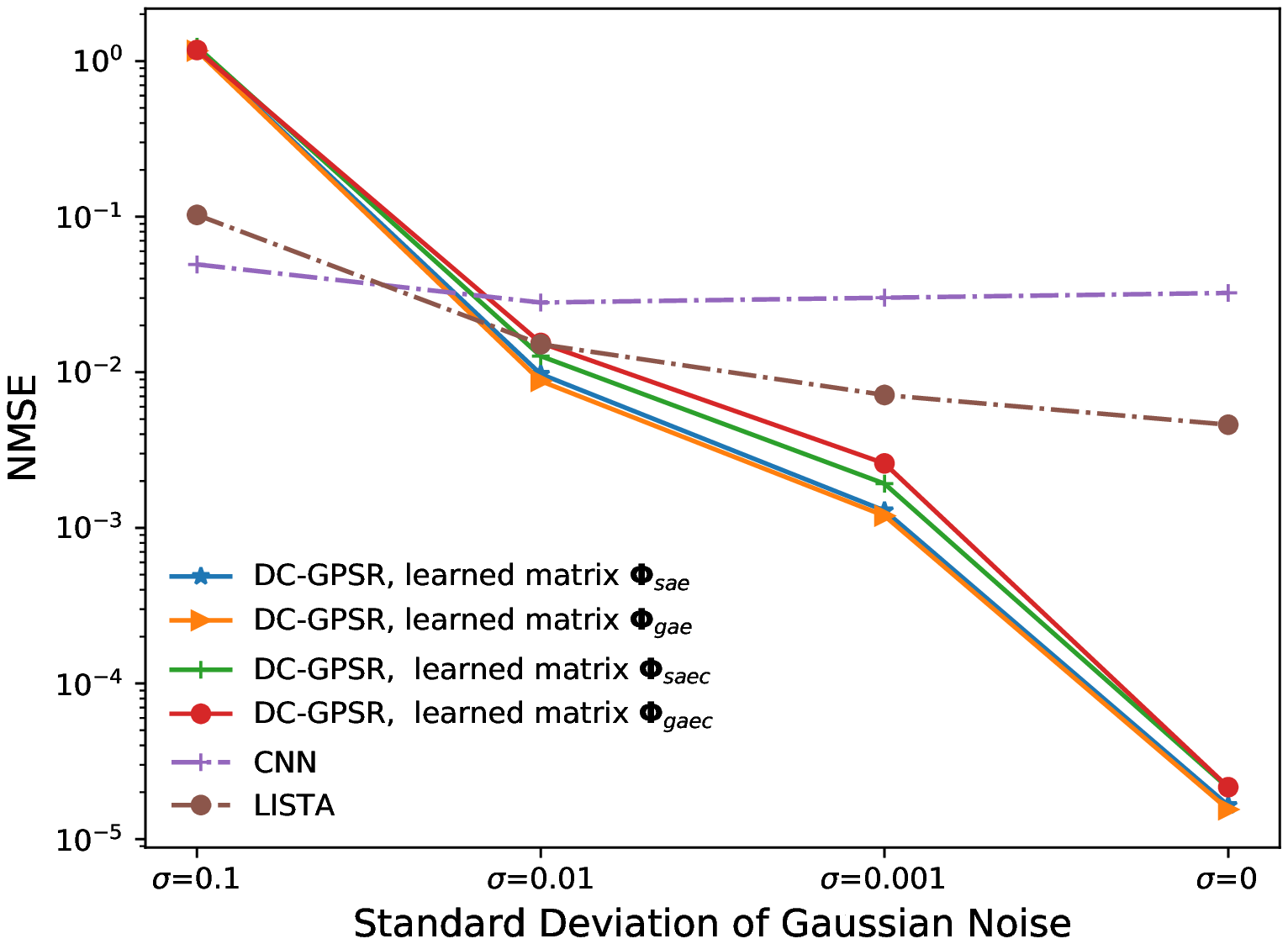}}
	\caption{Reconstruction NMSE for different noise levels}
	\label{exp_diag}
	\end{figure*}

%

\section{Conclusions}
	We proposed a data-driven approach to acquire measurement matrices.
	Specifically, we proposed a generic model-based autoencoder framework called deep unfolding $BP\textit{-}AE$ that consists of a linear dimensional-reduction encoder and a decoder that stacks the iterations to mimic basis pursuit sparse reconstructions.
	Under this framework, we constructed two model-based autoencoders, $BP\textit{-}sAE$ and $BP\textit{-}gAE$, and two extension models, $BP\textit{-}sAEcat$ and $BP\textit{-}gAEcat$.
	The proposed autoencoder models are parameterized by the measurement matrix so that they can be trained to optimize the measurement matrices given the sparse beamspace channel dataset.
	We provided numerical demonstrations to show that the learned data-driven measurement matrices can work well with general compressive sensing reconstruction algorithms, such as the linear programming, GPSR and DC-GPSR algorithms.
	 The learned data-driven matrices can attain higher achievable rates for CSI acquisitions than conventional random matrices, because more accurate reconstructions can be accomplished using fewer measurements.
	Therefore, the learned data-driven measurement matrices can be applied to design pilots or compress CSI feedback to reduce the overheads of downlink CSI acquisitions.
\bb{While we focused on the application of massive MIMO channel estimation, the proposed data-driven measurement matrix design method can be generalized in a straight-forward way to various wireless communication applications that use sparse reconstruction algorithms. On one hand, we demonstrated that the domain knowledge from traditional algorithms can be used to develop model-based deep learning methods. On the other hand, we demonstrated that the data-driven element can be customized and successfully incorporated into classical model-based algorithm pipelines to develop hybrid data-driven approaches. For future work,  we will consider to unfold alternative algorithms, such as the matching pursuit algorithm [25], to optimize the measurement matrix and the pilot matrix. We will also extend the proposed methods to mmWave massive MIMO systems where we may also consider a complex-valued measurement matrix under a hybrid analog-digital architecture.}

\appendix
\subsection{Experimental Results of DeepMIMO Dataset}	
	To validate the proposed autoencoders using a more realistic dataset, we present experimental results produced by the DeepMIMO dataset \cite{alkhateeb2019deepMIMO} and compare it to the performance produced by the dataset generated from channel model described by (1).
	The channel vectors were generated using the parameters summarized in Table VI, where the ‘O1’ ray-tracing scenario represents an outdoor signal propagation setup produced by the 3D ray-tracing simulator Wireless InSite. 
	The generated spatial-domain channel vectors were transformed into the beamspace channel vectors with the sparsity $S=16$, which is consistent with the generated dataset used in Section IV. It is worth noting that the DeepMIMO dataset has the DoAs that are associated with the signal propagation environment in the simulated scenario and are affected by the geometry distributions of surrounding streets and buildings, while the DoAs of our generated dataset using the channel model in (1) are uniformly distributed within the range of $[-\pi/2, \pi/2]$. 
	We choose the compressed dimension $M=32$ and repeat the same training and evaluation processes as in Section IV.A and Section IV.B. 
	The reconstruction NMSE and accurate-reconstruction percentage are summarized in Table VII.
	To interpret the improved performance produced by the DeepMIMO dataset, we observe that the channel vectors generated by DeepMIMO have a noticeable feature, i.e., the DoAs tend to concentrate into a limited range due to the geometry setting of the DeepMIMO simulation environment. 
	As a result, the support of the dataset tends to concentrate in a limited area. 
	This structural feature of the generated dataset is beneficial for the deep networks to exploit and leads to a significantly improved reconstruction result.
	On the contrary, our generated dataset, using the channel model in (1), was set to have uniformly distributed AoDs. Theoretically, the uniform distribution is one of the most challenging statistics for deep networks to exploit.

\begin{table*}[!t]
\centering
\caption {DeepMIMO Dataset Parameters} 
\label{Deepmimo_data} 
\begin{tabular}{|l|l||l|l|}
\hline
Parameters  & Values & Parameters  & Values\\
\hline
Ray-tracing scenario  & `O$1$' & System bandwidth & $0.5$ GHz\\
\hline 
 Activate BS & BS $4$   & Number of OFDM subcarriers & 1024\\ 
\hline
Activate users & Rows from R$1200$ to R$1500$ &  OFDM sampling factor & 1  \\ 
\hline
Number of BS antennas & $M_x=256$ &  OFDM limit & 1  \\
\hline
Antenna spacing (in wavelength) & $0.5$ & Number of channel paths & 3\\
\hline
\end{tabular}
\end{table*}

\begin{table*}[t!]
  \begin{center}
  \captionsetup{justification=centering}
    \caption{Reconstruction NMSE (Left) and Accurate-Reconstruction Percentage (Right)}
    \label{deepmimo_result}
    \begin{tabular}{|c|c|c|c|c||c|c|c|c|} 
    \hline
    \diagbox{Dataset}{Matrices} &  $\mathbf \Phi_{sae}$ & $\mathbf \Phi_{gae}$  & $\mathbf \Phi_{saec}$  & $\mathbf \Phi_{gaec}$ & $\mathbf \Phi_{sae}$ & $\mathbf \Phi_{gae}$  & $\mathbf \Phi_{saec}$  & $\mathbf \Phi_{gaec}$ \\
    \hline
     DeepMIMO dataset & $1.8 \times 10^{-3}$ & $8.5 \times 10^{-4}$ & $2.1 \times 10^{-3}$ & $1.6 \times 10^{-3}$  & 82.1\% & 87.6\% & 72.4\% & 81.1\%\\
    \hline
     Model (1) dataset & $1.4 \times 10^{-2}$  & $1.4 \times 10^{-2}$  & $6.2 \times 10^{-3}$  & $4.5 \times 10^{-3}$ & 44.2\% & 53.8\% & 61.9\% & 60.1\% \\
    \hline
    \end{tabular}
  \end{center}
\end{table*}

\bibliographystyle{IEEEtran}
\bibliography{IEEEabrv,thesis_bib}

\end{document}